\documentclass[12pt]{article}
\usepackage{amsbsy,amsmath}
\usepackage{amsfonts}
\usepackage{amssymb}
\usepackage{amscd}
\usepackage{bbm}
\usepackage{fancybox}
\usepackage{cite}
\usepackage{amsmath,amsfonts,amsbsy}
\usepackage{pstricks,pst-node}
\usepackage[small,bf,hang]{caption2}
\usepackage{graphicx}
\usepackage{epsfig}
\usepackage{psfrag}
\usepackage{comment}

\usepackage[mathscr]{euscript}

\usepackage{float}

\psset{unit=1.3cm,linewidth=.5pt,radius=.2}  

\usepackage{multirow}                     
\usepackage{float}                          
\usepackage{lscape}                         
\usepackage{bm}

\newcommand{\be}{\begin{equation}}
\newcommand{\ee}{\end{equation}}
\newcommand{\bea}{\begin{eqnarray}}
\newcommand{\eea}{\end{eqnarray}}
\newcommand{\ba}{\begin{array}}
\newcommand{\ea}{\end{array}}

\def\HH{\mathcal{H}}
\def\LL{\mathcal{L}}
\def\NN{\mathcal{N}}

\def\EE{\mathcal{E}}

\font\mybb=msbm10 at 12pt
\def\bb#1{\hbox{\mybb#1}}

\makeatletter \@addtoreset{equation}{section} \makeatother

\def\slashchar#1{\setbox0=\hbox{$#1$}           
   \dimen0=\wd0                                 
   \setbox1=\hbox{/} \dimen1=\wd1               
   \ifdim\dimen0>\dimen1                        
      \rlap{\hbox to \dimen0{\hfil/\hfil}}      
      #1                                        
   \else                                        
      \rlap{\hbox to \dimen1{\hfil$#1$\hfil}}   
      /                                         
   \fi}

\begin{document}

\begin{titlepage}

\begin{center}

\vskip 1.5cm

{\Large \bf Black holes and causal nonlinear electrodynamics}

\vskip 1cm

{\bf Jorge G.~Russo\,${}^{a,b}$ and  Paul K.~Townsend\,${}^c$} \\

\vskip 25pt

{\em $^a$  \hskip -.1truecm
\em Instituci\'o Catalana de Recerca i Estudis Avan\c{c}ats (ICREA),\\
Pg. Lluis Companys, 23, 08010 Barcelona, Spain.
 \vskip 5pt }

\vskip .4truecm

{\em $^b$  \hskip -.1truecm
\em Departament de F\' \i sica Qu\` antica i Astrof\'\i sica and Institut de Ci\`encies del Cosmos,\\ 
Universitat de Barcelona, Mart\'i Franqu\`es, 1, 08028
Barcelona, Spain.
 \vskip 5pt }
 
 \vskip .4truecm

{\em $^c$ \hskip -.1truecm
\em  Department of Applied Mathematics and Theoretical Physics,\\ Centre for Mathematical Sciences, University of Cambridge,\\
Wilberforce Road, Cambridge, CB3 0WA, U.K.\vskip 5pt }

\hskip 1cm

\noindent {\it e-mail:}  {\texttt jorge.russo@icrea.cat, pkt10@cam.ac.uk}

\end{center}

\vskip 0.5cm
\begin{center} {\bf ABSTRACT}\\[3ex]
\end{center}

For generic theories of nonlinear electrodynamics (NLED) we investigate the implications of (a)causality on  spherically-symmetric solutions of the Einstein-NLED equations that are asymptotic to a Reissner–Nordstr\" om (RN) spacetime. 
Equal-charge dyonic RN black holes are shown to be exact, but unstable, solutions of (acausal) ``Born-type'' theories. 
For {\sl all causal theories} it is shown that the metric is singular at the centre of symmetry and that it has at most two Killing horizons, implying at most three ``phases'': RN-like or S(chwarzschild)-like black holes, and naked timelike singularities. For extreme RN-like black holes, including dyons, we give simple proofs of monotonicity conditions that imply a reduction of mass and entropy due to NLED interactions. 
We find that causality allows four qualitatively different phase-diagrams. One of the two with finite electromagnetic energy $\EE_{\rm em}$ is the previously studied Born-Infeld-type, for which the zero-entropy limit of a ``small-charge''  S-like black hole is a naked timelike singularity of mass $M=\EE_{\rm em}$; we show that the spacetime geometry of this ``Born particle'' is that of the Bariola-Vilenkin global monopole.

\vfill

\end{titlepage}
\tableofcontents

\section{Introduction}
\setcounter{equation}{0}

The Einstein-Maxwell field equations have as solutions the three-parameter
family of Reissner-Nordstr\" om (RN) charged black holes, with a spacetime 
metric that is asymptotically-flat and spherically-symmetric. The parameters are the black-hole mass $M$ and its electric and magnetic charges $(q_e,q_m)$, although the RN metric depends only on the two parameters $M$ and 
\be\label{Q1}
Q= \sqrt{q_e^2 + q_m^2}\, . 
\ee
An important property of the Einstein-Maxwell field equations of relevance to the classical physics of black holes is that they are second-order partial differential equations, and this remains true if Maxwell electrodynamics is replaced by any theory of nonlinear electrodynamics (NLED) defined by a scalar Lagrangian function $L(F)$ of the Faraday field strength 2-form $F=dA$ but not derivatives of $F$. It is therefore natural to consider black holes as solutions of the more general Einstein-NLED field equations.  

Here we consider only NLED theories with a weak-field expansion and a conformal weak-field limit. Simple examples are Born's original 1933 NLED theory \cite{Born:1933qff} and the subsequent Born-Infeld modification of it \cite{Born:1933pep}. In both cases the Lagrangian depends on a constant $T$ with dimensions of energy density (this is the ``Born tension'', which is the square of the ``Born constant''). The $T\to\infty$ limit for fixed fields is a weak-field limit to Maxwell, and the weak-field expansion is a series expansion of $L(F)$ in powers of $1/T$. More generally we can expect the interactions in $L(F)$ to depend on a number of dimensionless parameters  and a single parameter $T$ that sets the scale for field-energy densities at which the interactions first become significant. 

A special subclass of NLED theories is composed of those with a Hamiltonian invariant under a $U(1)$ electromagnetic duality group. These are ``self-dual''; the free-field Maxwell theory is an example, as is the Born-Infeld (BI) theory.  
Our derivation of static charged black-hole solutions of the Einstein-NLED equations makes it manifest that for self-dual NLED the spacetime metric can depend on the charges $(q_e,q_m)$ only through its dependence on $Q$, and more generally that duality symmetries of the Hamiltonian imply corresponding symmetries in the $(q_e,q_m)$ parameter space of the spacetime metric. This result is perhaps not surprising but we know of no previous proof of it.  Another example is the $Z_2$-duality invariance of the original Born theory, which implies isometric electric and magnetic black hole spacetimes.  

One of the expected effects of replacing the free-field Maxwell theory by some interacting NLED theory is that asymptotically-flat spherically-symmetric black-hole solutions of the Einstein-NLED equations will be only asymptotically isometric to the RN solutions of the Einstein-Maxwell equations. The mass $M$ and charge $Q$ of the RN spacetime determine two distinct length scales; these are $GM$ and $\sqrt{G}\, Q$, where $G$ is Newton's constant. For any non-conformal NLED theory the Born tension $T$ determines an additional ``gravitational-Born'' length scale:
\be\label{gBdef}
\ell_{gB} \sim 1/\sqrt{GT}\, .
\ee
Notice that this is zero in the $T\to\infty$ limit, which is a weak-field limit. We can expect (and it is true) that the existence of an event horizon requires $GM \gtrsim\sqrt{G}\, Q $.
This suggests that we should expect significant modifications to the RN metric when
\be\label{strong}
\ell_{gB} \gtrsim \sqrt{G}\, Q \, ,  
\ee  
which can be viewed as a ``small charge'' condition. 

Despite the expectation of deviations from the RN metric, a natural question is whether there are NLED theories for which the RN metric is an exact solution for some choice of charges $(q_e,q_m)$. The answer is no for causal theories but yes for some acausal theories; the RN equal-charge ($q_e=q_m$) dyonic black hole is shown to be an exact solution for the entire class of ``Born-type'' theories with Lagrangian function $L(S)$, all of which are acausal for sufficiently strong fields
\cite{Schellstede:2016zue}. We use the prototype example of Born's original theory to show how acausality leads to instability against some perturbations of the electric and magnetic fields that break the spherical symmetry.

Generic NLED theories are acausal because they allow superluminal propagation of shock-wave discontinuities in smooth electromagnetic backgrounds \cite{Boillat:1970gw,Plebanski:1970zz}  or (equivalently)
of perturbations of stationary homogeneous backgrounds, which can be viewed as anisotropic (and generically birefringent) optical media \cite{Bialynicki-Birula:1984daz}.
The necessary and sufficient conditions on $L(F)$ for this not to be possible, i.e. for causality, were found in \cite{Schellstede:2016zue}. They comprise standard convexity conditions which are necessary for ``weak-field" causality  but also an additional condition needed for ``strong-field" causality, which we have rederived, discussed and applied in recent works \cite{Russo:2024kto,Russo:2024llm,Russo:2024xnh,Russo:2024ptw,Russo:2025fuc}.
It is important to appreciate that both ``weak-field" and ``strong-field" causality must be satisfied for all values of the fields and, as mentioned, these are necessary and sufficient conditions for causality.
The ``weak/strong" terminology here arises because the weak-field causality condition can be violated by arbitrarily weak fields, whereas the violation of the strong-field causality condition requires some field to be strong.

Our principal aim here is to deduce implications of causality for spherically-symmetric charged black-hole solutions of the Einstein-NLED equations; such solutions are necessarily static by Birkhoff's theorem (see e.g. \cite{Bronnikov:2019fgh}).
 There are several issues of interest for which causality plays a role in restricting possibilities. One is whether the central spacetime singularity of a charged static black-hole solution of the Einstein-NLED equations can be absent for some choice of NLED theory \cite{deOliveira:1994in,Bronnikov:2000vy,Bronnikov:2022ofk,Bokulic:2022cyk,Abe:2025vdj,Hale:2025ezt,Croney:2025cue,Babaei-Aghbolagh:2025tim,Hale:2025urg}. 
A restriction on this choice is inherent in the Hawking-Penrose theorem that, for a stress-energy tensor satisfying the Strong Energy Condition (SEC), a spacetime singularity is inevitable once a trapped surface has appeared \cite{Hawking:1970zqf}. Thus, ``regular'' (singularity-free) charged black hole solutions require an SEC-violating NLED theory, but the SEC is implied by the strong-field causality condition mentioned above \cite{Russo:2024xnh}. This rules out regular asymptotically-flat black-holes.

However, it is still desirable to have a detailed understanding of why a spacetime singularity is 
unavoidable in the Einstein-NLED context. Here 
we clarify and complete some partial arguments of Bronnikov \cite{Bronnikov:2000vy,Bronnikov:2022ofk} that there are no regular charged  spherically-symmetric black holes for any causal NLED theory, and it is not difficult to see that the inclusion of a cosmological constant cannot eliminate a central singularity. In addition, we also give a direct proof 
based on the fact that the Ricci-squared scalar is strictly positive, and singular at the centre of symmetry, as a result of the strong-field causality condition on generic NLED theories. 

Only radial electric/magnetic fields, depending only on a radial coordinate $r$, are compatible with spherical symmetry and time-independence, which means that the NLED Hamiltonian function $H$ becomes what we shall call a ``black-hole'' function: ${\rm H}(r)$. Many features of spherically-symmetric charged black holes depend crucially on the behaviour of this function near the central singularity: $r=0$ in Schwarzschild coordinates. For most cases discussed in the literature the leading term as $r\to0$ is a power of $r$: 
\be\label{powerlaw}
{\rm H}(r) \sim r^{-4\nu} \qquad (r\to 0)\,  
\ee
for some constant $\nu$. For RN black holes $\nu=1$ because the electric/magnetic fields diverge as $1/r^2$ and $H$ is a quadratic function of these fields. Although not all charged black holes of causal Einstein-NLED theories have a black-hole function with a power-law behaviour as $r\to0$, only four  values or ranges of values of $\nu$ are ultimately 
relevant and the distinctions between them are generally 
applicable, as we explain in section \ref{sec:ADM}.


The black-hole function ${\rm H}(r)$ determines two other functions that are important to properties of spherically-symmetric black-hole spacetimes. One is a function 
$\EE(r)$, which gives the electromagnetic energy outside a sphere of radius $r$. The total electromagnetic energy is therefore 
\be
\EE_{\rm em} := \lim_{r\to0} \EE(r)
\ee 
when this limit exists; otherwise it is infinite (as it is for RN black holes).  
The other is an  ``effective charge'' function $Q_{\rm eff}(r)$, which is asymptotic to $Q$ as $r\to\infty$ (assuming that Maxwell is the weak-field limit). It can be defined by its relation to $\EE(r)$: 
\be
2\pi Q^2_{\rm eff}(r) := r\EE(r)\, . 
\ee
The importance of these two functions is that they are restricted in a very simple way by causality. It was shown by Hale et al, that the SEC (which, as mentioned above, is implied by strong-field causality) implies that $\mathcal{E}(r)$ is convex \cite{Hale:2025ezt} (in fact, it is strictly convex) and we show here that weak-field causality requires $Q^2_{\rm eff}(r)$ to be concave. The combination of these two convexity/concavity constraints imposes both lower and upper limits on the power-index $\nu$: 
\be\label{nurange-intro}
\tfrac12 \le \nu\le 1 \, . 
\ee
The value of $\nu$ also determines (for any causal NLED) whether the electric field is finite at $r=0$ (it is iff $\nu=\frac12$) and whether the total electromagnetic energy is finite (it is iff $\frac12 \le \nu<\frac34$). 

An assumption about the $r\to0$ behaviour of $\mathcal{E}(r)$ near $r=0$ was made in \cite{Hale:2025ezt}, and it agrees with what we find from \eqref{powerlaw} for $\nu=\frac12$, which includes Born-Infeld. However, it disagrees with what we find for $\frac12 < \nu<\frac34$; i.e. the cases (not included in the analysis of \cite{Hale:2025ezt}) for which the electromagnetic energy is finite but the electric field diverges as $r\to 0$. This can be seen by consideration of a simple ``minimal'' family of causal self-dual NLED theories for which the black-hole function ${\rm H}(r)$ has the power-law behaviour of \eqref{powerlaw} with a power-index that varies over the entire range of \eqref{nurange-intro}.  

A major aim of this paper is to explore the implications of causality for the global structure of Einstein-NLED charged black-holes, assuming spherical symmetry and asymptotic flatness. For Maxwell we have the RN family of spacetimes with two Killing horizons, an event horizon and an interior Cauchy horizon. For generic NLED theories there can be more than two Killing horizons \cite{Nojiri:2017kex,Gao:2021kvr} but two is the maximum number for causal NLED theories. A proof based on convexity of $\EE(r)$ was given in \cite{Hale:2025ezt} but with an implicit restriction to BI-type; here we extend this argument to all black hole solutions for any causal NLED theory. Whenever there are precisely two Killing horizons they can be identified as the event horizon and an (interior) Cauchy horizon. For generic NLED theories we shall say that these are ``RN-like'' black holes.

For Born-Infeld, it was shown long ago by Oliveira \cite{deOliveira:1994in} that all black holes with a charge $Q$ less than a critical value $Q_{\rm cr}$ (with $\sqrt{G}Q_{\rm cr} \sim \ell_{gB}$) have a Killing horizon but no Cauchy horizon; following \cite{Hale:2025ezt} we shall call them ``S-like'' black holes because their global structure is similar to that of uncharged Schwarzschild black holes. An important observation in \cite{Hale:2025ezt} is that S-like black holes holes are possible only when the electromagnetic energy is finite, but the additional claim that ``any causal theory of nonlinear electrodynamics that regularizes the point charge self-energy also eliminates Cauchy horizons for weakly charged black holes'' is too strong; it applies only to BI-type black holes. We give examples of causal NLED theories with finite electromagnetic energy for which RN-like black holes exist for arbitrarily small charge. 

As for RN black holes, coincidence of the Cauchy and event horizons leads to a degenerate event horizon (zero surface gravity and hence zero Hawking temperature in the quantum theory), and this occurs when the mass has its lowest (extreme) value (at fixed $Q$) compatible with the existence of an event horizon: $M= M_{\rm ext}$. 
Let $\mu(Q)$ be the extreme black hole mass-to-charge ratio
normalised such that $\mu=1$ for Maxwell, which implies that $\mu(Q)\to 1$ as $Q\to\infty$ for all other cases. It was shown by Abe et al. \cite{Abe:2025vdj}, using a Lagrangian formulation and an initial restriction to zero electric charge, that $\mu(Q)$ is a monotonic increasing function for causal NLED theories. We give here an alternative proof, valid for any $q_e/q_m$ ratio, using only weak-field causality.  

We also prove a similar monotonicity property for the ratio
$r_h(Q)/Q$ where $r_h$ is the event-horizon radius of an extreme black hole. However, this result relies on both 
weak-field causality {\sl and} strong-field causality. 
It implies that causal NLED interactions reduce (at fixed $Q$) the event-horizon area, and hence the entropy in the quantum theory.  

For any NLED theory, and any spherically-symmetric solution of the Einstein-NLED equations, with charges $(q_e,q_m)$, the black-hole function ${\rm H}(r)$ can be found on a case-by-case basis if the explicit NLED Hamiltonian function $H$ is known. One can do better for self-dual theories because the spherical-symmetry restriction implies that $H$ is a function of a single duality-invariant scalar variable $s$ (the free-field energy density), and $2s= Q^2/r^4$. In these cases the form of $H(s)$ as $s\to\infty$ directly determines the form of the function ${\rm H}(r)$ as $r\to0$, which means that the power-index $\nu$ becomes a characteristic of the self-dual NLED theory; e.g. $\nu=\frac12$ for Born-Infeld. Moreover, causality conditions for self-dual NLED reduce to simple conditions on $H(s)$ \cite{Russo:2024ptw}, which lead directly to \eqref{nurange-intro}. 

Almost all causal NLED theories for which an explicit Lagrangian or Hamiltonian is known are self-dual, and for this
reason most of the examples we consider for illustrative purposes are self-dual. To compensate for this bias, we investigate black hole solutions for a simple NLED theory with an explicit Hamiltonian that has no electromagnetic duality invariance (neither $U(1)$ nor $Z_2$), and is known to be causal \cite{Russo:2024kto}. For this non-self-dual example the electric and magnetic black holes have different values of the power-index $\nu$, and the behaviour of the black-hole function near the spacetime singularity changes discontinuously at zero magnetic charge. Nevertheless, for any choice of the $q_e/q_m$ ratio, there is a definite black-hole function ${\rm H}(r)$ with a definite value of $\nu$.  

Since causality allows a maximum of two Killing horizons, it also allows a maximum of three kinds of spherically symmetric spacetime with (positive) ADM mass $M$ and charge $Q$: RN-like black holes, S-like black holes and spacetimes with a naked timelike singularity. Depending on the parameters $(M,Q)$, and on the NLED theory, one of these three possibilities is realized; we call them ``phases'', and their distribution in the $(M,Q)$ parameter space is then a ``phase diagram''. 

Although we have used the power-index $\nu$ as a way of characterising charged black hole solutions, many different values of $\nu$ lead to qualitatively indistinguishable 
phase diagrams. In fact, there are only four qualitatively distinct ``types''of phase diagram, corresponding to four ``types'' of charged black hole:

\begin{itemize}

\item {\sl Maxwell-type} : $\quad \nu=1$. Infinite electromagnetic energy.
This includes a novel  (and previously unappreciated)  class of causal self-dual NLED 
theories for which ModMax \cite{Bandos:2020jsw} is a conformal strong-field limit. A simple example is given in section \ref{sec:global}.

\item {\sl Intermediate-type I} :  $\quad \frac34 \le \nu < 1$.
Infinite electromagnetic energy, as for $\nu=1$, but with
$\mu(Q)\to0$ as $Q\to 0$ (as for Intermediate-type II)

\item {\sl Intermediate-type II} : $\quad \frac12 < \nu < \frac34$. Finite electromagnetic energy, as for $\nu=\frac12$, but without a critical charge.

\item {\sl Born-Infeld-type} : $\quad \nu=\frac12$. Finite electromagnetic energy. The phase diagram for this case was given in \cite{Hale:2025ezt} (using a different parametrisation from the one used here).  

\end{itemize}

Of these four types of phase diagrams, the BI-type 
has the most novel features; in particular, the transition, for subcritical charge, from an S-like black hole to a naked timelike singularity at non-zero mass \cite{deOliveira:1994in,Hale:2025ezt}. In fact, the mass at this transition point is the electromagnetic energy: $M=\EE_{\rm em}$. Since this transition point can be interpreted as a zero-entropy limit  for a ``small-charge'' S-like black hole, it is potentially a (gravitational) realisation of Born's idea that the mass of a charged point particle could, in a nonlinear theory of electrodynamics, equal the energy in its electromagnetic fields \cite{Born:1933qff}. Here we show that the spacetime geometry of this putative ``Born particle'' is the same as the geometry of the Barriola-Vilenkin global point monopole \cite{Barriola:1989hx}. 

\section{NLED in general spacetimes}\label{sec:ADM}
\setcounter{equation}{0}

The general  NLED with a weak-field limit has a Lagrangian density $\LL$ that is a function 
of two independent Lorentz scalars quadratic in the field-strength 2-form $F=dA$. 
If we allow for a generic spacetime with metric $g$, then 
\be \label{Pleb}
\LL = \sqrt{|g|}\,  L(S,P)\, , 
\ee
where $|g| = -\det g$,  and $L$ is a scalar function of the (pseudo)scalars 
\be
S = -\frac14 g^{\mu\rho} g^{\nu\sigma} F_{\mu\nu} F_{\rho\sigma}\, , \qquad 
P = - \frac{1}{8\sqrt{|g|}} \varepsilon^{\mu\nu\rho\sigma} F_{\mu\nu} F_{\rho\sigma}\, . 
\ee
Maxwell electrodynamics corresponds to $L=S$ in units for which $c=1$, where $c$ is the {\it in vacuo} 
speed of light. The NLED stress-energy tensor can be found from the Hilbert formula 
\be\label{Hilbert}
T_{\mu\nu} = -\frac{2}{\sqrt{|g|}} \frac{\partial \LL}{\partial g^{\mu\nu}}\, . 
\ee
This yields 
\be
\label{entensor}
T_{\mu\nu} = L_S T^{\rm Max}_{\mu\nu} - \left(SL_S + PL_P -L\right)g_{\mu\nu}\, , 
\ee
where the Maxwell stress-energy tensor is 
\be
T^{\rm Max}_{\mu\nu} = g^{\rho\sigma}F_{\mu\rho}F_{\nu\sigma} + Sg_{\mu\nu}\, .
\ee
Since the Maxwell stress-energy tensor is traceless (with respect to the metric $g$) we have 
\be\label{Theta<0}
\Theta := g^{\mu\nu} T_{\mu\nu} = -4(SL_S+ PL_P -L)  \le0 \, , 
\ee
where the inequality follows from convexity of the function $\LL(S,P)$ (required for weak-field causality)
and the assumption of zero vacuum energy (required for an asymptotically-flat spacetime). This is because $SL_S+ PL_P \ge L$ for any convex function $L(S,P)$ such that $L(0,0)=0$. Recall that one definition of a convex function is that every plane tangent to its graph must be nowhere above it.

\subsection{ADM coordinates and the Hamiltonian}

To pass to the Hamiltonian formulation it is useful to choose coordinates $x^\mu=(t,x^i)$ for which the spacetime metric takes the following (ADM) form:
\be\label{ADM}
ds^2(g) = - \NN^2 dt^2 + h_{ij} \left(dx^i + u^i dt\right)\left(dx^j + u^j dt\right)\,  \qquad 
\left(\Rightarrow\ |g|= \NN^2\det h \right). 
\ee
In this case the (pseudo)scalars $(S,P)$ are 
\be\label{S&P}
S= \frac{1}{2} \left\{ \frac{1}{\NN^2} |E-u\times B|^2 - \frac{1}{\det h} |B|^2\right\} \, , \qquad P = \frac{1}{\NN \sqrt{\det h}} E_iB^i\, , 
\ee
where 
\be 
E_i := F_{i0} = \partial_t A_t - \partial_t A_i\, , \qquad 
B^i := \frac12\varepsilon^{ijk}F_{jk} \equiv \varepsilon^{ijk}\partial_j A_k\, ,  
\ee
and 
\be 
(u\times B)_i := \varepsilon_{ijk} \, u^jB^k \, , \qquad \left(\varepsilon^{ijk} \varepsilon_{lmn} = 6\,\delta_{(l}^i \delta_m^j \delta_{n)}^k\right).
\ee
The norm $|..|$ is taken using the 3-metric $h$; e.g. 
\be
|E|^2 = h^{ij} E_iE_j \, , \qquad |B|^2 = h_{ij}B^iB^j \, .
\ee

For this choice of spacetime coordinates, the dielectric displacement field is
\be\label{DD}
D^i := \frac{\partial \LL}{\partial E_i} 
= L_S\frac{\sqrt{\det h}}{\NN}\, h^{ij}(E-u\times B)_j + L_P B^i\, . 
\ee
Solving this equation for $E$ as a function of $D$ allows us to find the 
Hamiltonian density $\HH(D,B)$ by Legendre transform of $\LL(E,B)$. The solution will be unique whenever $\LL(E,B)$ is a strictly convex function of $E$. This is equivalent to strict convexity of the function $L(S,P)$ combined with the condition $L_S>0$ \cite{Bandos:2021rqy}, which are necessary conditions for causality of any NLED with a weak-field limit \cite{Schellstede:2016zue}, and sufficient for {\sl weak-field} causality \cite{Russo:2024kto}.

As it is not always possible to find $\HH(D,B)$ explicitly from $\LL(E,B)$, a useful alternative starting point is the `phase-space' Lagrangian density: 
\be\label{psLag}
\LL' = E_iD^i - \HH(D,B)\, , 
\ee
where $D$ is now an independent field and $\HH$ is the Hamiltonian density.
The integral of $\HH$ over any volume $V$ at fixed time $t$ is the 
electromagnetic energy in $V$:
\be\label{emen}
\mathcal{E}_{\rm em}(V) = \int_V d^3x \, \HH(D,B)\, . 
\ee
The field equations that follow from $\LL'$ imply that 
\be
\dot \HH =  -\partial_i(E\times H)^i \, ,  \qquad 
H_i := \frac{\partial \HH }{\partial B^i}\, . 
\ee
For any NLED that is Lorentz invariant, we have\footnote{In local coordinates for which $h_{ij}=\delta_{ij}$, this is the condition for symmetry of the stress-energy tensor, which is necessary and sufficient for Lorentz invariance given time and space translation invariance and rotation invariance \cite{Bialynicki-Birula:1984daz,Mezincescu:2023zny}.}
\be\label{LLI}
(E\times H)^i =  h^{ij} p_j  \, \qquad p_j:= (D\times B)_j\, , 
\ee
where $p_i$ is the 3-momentum density; this is zero for static electromagnetic fields, and therefore $\mathcal{E}_{\rm em}(V)$ is time-independent for static fields. Notice that both $D$ and $B$ are divergence-free since $\partial_iB^i=0$ is an identity and $\partial_iD^i=0$ is the constraint imposed by the Lagrange multiplier $A_t$ (in the absence of electric charges). Variation of $D$ in $\LL'$ yields 
\be\label{defE}
E_i = \frac{\partial \HH}{\partial D^i} \, . 
\ee
This equation will uniquely determine $D$ as a function of $E$ if $\HH$ is a strictly convex function of $D$, which is required for causality of any non-conformal NLED with a weak-field expansion\footnote{If the convexity is not strict then there will be Lagrangian constraints and no weak-field limit.}. Thus, $D$ is an auxiliary field in this context. 

For the general spacetime metric of \eqref{ADM} we have
\be
\HH(D,B) = \NN\sqrt{\det h} \, H(D,B) - u^i p_i\, , 
\ee
where $H(D,B)$ may be expressed, assuming rotation invariance, as a function 
$H({\rm x},{\rm y},{\rm z})$ of the following three 3-space rotation scalars: 
\be
{\rm x} = \frac{1}{2\det h}  h_{ij} D^iD^j\, , \qquad {\rm y}= \frac{1}{2\det h} h_{ij} B^iB^j\, , 
\qquad {\rm z} = \frac{1}{\det h} h_{ij} D^iB^j\, . 
\ee
We may now rewrite \eqref{defE} as 
\be\label{defE'}
E_i = \left(\frac{\NN}{\sqrt{\det h}}\right) h_{ij}\left(H_{\rm x} D^j + H_{\rm z} B^j\right)\, . 
\ee
Using this to eliminate $D$ from $\LL'$ yields a Lagrangian density $\LL(E,B)$ but this will be expressible in the `Plebanski' form of \eqref{Pleb} only if the (local) Lorentz invariance condition \eqref{LLI} is satisfied; for Hamiltonian functions $H({\rm x},{\rm y},{\rm z})$ this condition is 
equivalent to \cite{Bialynicki-Birula:1984daz,Mezincescu:2023zny}
\be\label{LIH}
H_{\rm x}H_{\rm y} - H_{\rm z}^2 =1\, . 
\ee
If this condition is satisfied and if $\HH$ is a strictly convex function of $D$ then 
$\HH(D,B)$ is the Legendre dual of the Lagrangian function $\LL(E,B)$ found by elimination of $D$ from $\LL'$. For example, the free-field Maxwell case is $H={\rm x} + {\rm y}$, and elimination of $D$ yields $\LL = \sqrt{|g|}\, S$, with $S$ given by \eqref{S&P}.

We remark that the Hilbert formula of \eqref{Hilbert} for the stress-energy tensor can also be used for $\LL'$, in which case it is equivalent to the relations
\be\label{stcpts}
\begin{aligned}
T_{tt} -2u^i T_{ti} + u^i u^j T_{ij} &= - \frac{\NN^2}{\sqrt{\det h}}  \frac{\partial \LL'}{\partial \NN} \ =  \NN^2 H\\
T_{ti} - u^j T_{ij} &= -\frac{\NN}{\sqrt{\det h}} \frac{\partial \LL'}{\partial u^i} \ = - \frac{\NN}{\sqrt{\det h}}\, p_i \\
T_{ij} &= -\frac{2}{\NN\sqrt{\det h}} \frac{\partial\LL'}{\partial h^{ij}} \ = 2 \frac{\partial H}{\partial h^{ij}} - h_{ij} H \, . 
\end{aligned}
\ee
For static electromagnetic fields on a static spacetime we have $u=0$ and $|D\times B|=0$, and all non-zero fields and metric components are time-independent. In this case (of relevance here) we have 
\be
H= \NN^{-2} T_{tt} = -T^t{}_t\, . 
\ee

\subsection{Self-dual NLED}\label{subsec:sd} 

For the subset of Lorentz-invariant NLED that are also $U(1)$-duality invariant (i.e. self-dual) the Lagrangian function $L$ must satisfy a particular nonlinear first-order differential equation \cite{Bialynicki-Birula:1984daz}. In terms of new scalar variables $(U,V)$ defined by \cite{Gibbons:1995cv,Gaillard:1997rt}
\be
S= V-U\, , \qquad P= \sqrt{4UV}\,  \qquad (V\ge U\ge0), 
\ee
this self-duality equation is $L_UL_V=-1$  and the general solution, given by Courant and Hilbert (CH) in \cite{C&H},  is 
\be
L(U,V) = \ell(\tau) - \frac{2U}{\dot\ell(\tau)} \, , \qquad 
\tau= V + \frac{U}{ \left[\dot\ell(\tau)\right]^2}\, , 
\ee
where $\ell(\tau)$ is a ``CH-function'', which has a Taylor expansion in powers of $\tau$ for self-dual NLED theories with a weak-field expansion:
\be
\ell(\tau) = e^\gamma \tau + \mathcal{O}(\tau^2) \qquad (\tau\ge0). 
\ee
The absence of a constant term implies zero vacuum energy. 
The linear term yields the conformal Maxwell/ModMax family with 
\be
\LL = e^\gamma V - e^{-\gamma} U \equiv (\cosh\gamma)S + (\sinh\gamma)\sqrt{S^2+P^2}\,,  
\ee
and causality requires $\gamma\ge0$ \cite{Bandos:2020jsw}. The higher powers of $\tau$ introduce the non-conformal interactions of the weak-field expansion, and NLED theories defined in this way are causal iff  \cite{Russo:2024llm} 
\be\label{causal-sd}
\dot\ell \ge1 \, \qquad \ddot\ell \ge0\, ,  
\ee
where the overdot indicates a derivative with respect to $\tau$. The first inequality is an equality only for Maxwell. The second inequality is an equality only for the Maxwell/ModMax family, which means that $\ell$ is a strictly convex function for any non-conformal causal self-dual NLED. 

Since $(U,V)=(0,S)$ when $B=0$, which implies $P=0$ and $S\ge0$, we have 
\be\label{elldef}
L(S,0) = \ell(S) \qquad (S\ge0),  
\ee
which can be used as a definition of the function $\ell$ for generic NLED theories. In this general (but $B=0$) context, the conditions \eqref{causal-sd} become conditions for weak-field causality; only for self-dual NLED do they also guarantee strong-field causality.

In the Hamiltonian formulation, self-duality can be made manifest by 
restricting $H$ to be a function of the two duality invariant variables\footnote{Notice that $p$ is the 3-space scalar constructed from the 3-momentum density $p_i$ defined in \eqref{LLI}.}
\be\label{s&p}
s := {\rm x}+{\rm y}\, , \qquad p := \sqrt{4{\rm xy}-{\rm z}^2} \equiv |D\times B|/\sqrt{\det h}\, . 
\ee
However, a function $H(s,p)$ will not define a Lorentz invariant NLED unless it satisfies a particular nonlinear first-order differential equation equivalent to \eqref{LIH}. In terms of new variables $(u,v)$ 
defined by 
\be
s= v+u\, , \qquad p= \sqrt{4uv}\, \qquad (v\ge u\ge 0)\, ,  
\ee
the general solution of this Lorentz-Invariance condition is 
\be\label{HamCH}
H(u,v) = \mathfrak{h}(\sigma) + \frac{2u}{ \mathfrak{h}'(\sigma)}\, , \qquad \sigma= v- \frac{u}{[\mathfrak{h}'(\sigma)]^2}\, , 
\ee
where $\mathfrak{h}$ is a ``Hamiltonian CH-function'', and the prime indicates a derivative with respect to the independent variable $\sigma$. The conditions for causality may now be expressed as the following conditions on this function \cite{Russo:2024ptw}:  
\be\label{hfrak-c}
0<\mathfrak{h}'(\sigma) \le 1\, , \qquad 
\mathfrak{h}''(\sigma) \le 0\, , \qquad 
\mathfrak{h}'(\sigma)  +2\sigma \mathfrak{h}''(\sigma)>0 \, .
\ee
Notice that for $p=0$ we have $(u,v)=(0,\sigma)$, and hence 
\be
\label{Hsss}
H(s) = \mathfrak{h}(s) \qquad (p=0)
\ee
This may be taken as a definition of $\mathfrak{h}$ for generic NLED, 
in which context the conditions \eqref{hfrak-c} become weak-field causality conditions that are sufficient for full causality {\sl only} for self-dual NLED.   

For any 
self-dual NLED theory,  $\ell(\tau)$ and $\mathfrak{h}(\sigma)$
are related as follows \cite{Russo:2024ptw,Russo:2025fuc}
\be
\label{relah}
\ell(\tau)=2\sigma \mathfrak{h}'(\sigma)-\mathfrak{h}(\sigma)\ ,\qquad \tau =\sigma\, \left[\mathfrak{h}'\right]^2\ ,
\ee
or, equivalently,
\be
\label{relahh}
\mathfrak{h}(\sigma)=2\tau \dot \ell(\tau)-\ell(\tau)\ ,
\qquad \sigma =\tau\, \left[\dot\ell(\tau)\right]^2\ .
\ee
These relations imply that 
\be\label{CHderivs}
\dot\ell(\tau)\mathfrak{h}'(\sigma)=1\, . 
\ee 

For any Lorentz invariant self-dual NLED with Hamiltonian  $H(s,p)$ its conformal weak-field limit is (if it exists) either Maxwell or ModMax \cite{Bandos:2020jsw}; the Maxwell/Modmax family has $\mathfrak{h}(\sigma)=e^{-\gamma}\sigma $ and
\be\label{ModMax}
H_{MM}(s,p)= (\cosh\gamma)s - (\sinh\gamma) \sqrt{s^2-p^2}\, , 
\ee
where $\gamma\ge0$ is a non-negative dimensionless coupling constant. Maxwell is the free-field $\gamma=0$ case (the 
$\gamma<0$ cases are acausal).

\section{Einstein-NLED equations and black holes}\label{sec:ADM}
\setcounter{equation}{0}

The Einstein field equation for the spacetime metric $g$ is 
\be\label{Einstein}
G_{\mu\nu} = (8\pi G)\, T_{\mu\nu}\, , 
\ee
where $G_{\mu\nu}$ and $T_{\mu\nu}$ are, respectively, the Einstein tensor and the NLED stress-energy tensor, and 
$G$ is Newton's constant. Here we restrict to spherically-symmetric\footnote{And hence static, by Birkhoff's theorem.} and asymptotically-flat spacetimes associated to charged (generically dyonic) black holes. In the context of the general spacetime metric of \eqref{ADM} this means that  
$u^i=0$ and both $\NN$ and $h_{ij}$ are time-independent. If we further assume spherical symmetry, and choose Schwarzschild radial coordinates, we arrive at a family of metrics parameterised by functions $\NN(r)$ and $h_{rr}(r)$:
\be
ds^2 =-\NN^2(r) dt^2 + h_{rr}(r) dr^2 + r^2(d\theta^2 + \sin^2\theta d\phi^2)\, .
\ee
The electromagnetic fields must also be static and compatible with spherical symmetry. This implies that the
only non-zero components of $(D,B)$ are the radial components $(D^r,B^r)$, which must be $t$-independent. They must also be $r$-independent by the divergence-free conditions, and scalar densities on the unit 2-sphere. Therefore
\be\label{radialDB}
D^r = q_e \sin\theta \, , \qquad  B^r= q_m\sin\theta\, ,   
\ee 
for constants $(q_e,q_m)$ that can be identified as the electric and magnetic charges. 
The 3-space scalars $({\rm x},{\rm y},{\rm z})$ are now\footnote{Since $4{\rm xy}-{\rm z}^2\equiv |D\times B|^2$, which must be zero for a static
solution of the NLED field equations, we have $4{\rm xy}={\rm z}^2$.}
\be\label{xyz} 
{\rm x}= \frac{q_e^2}{2r^4} \, , \qquad {\rm y}= \frac{q_m^2}{2r^4}\, , 
\qquad {\rm z} = \frac{q_e q_m}{r^4}\, . 
\ee

From \eqref{stcpts} we see that the only non-zero components of the NLED stress-energy tensor for static field configurations are
\be
T_{tt} = \NN^2 H\, , \qquad  T_{ij} = 2 \frac{\partial H}{\partial h^{ij}} - h_{ij} H \, .
\ee
The further restriction to spherical symmetry implies that only the diagonal components of $T_{ij}$ are 
non-zero, and because the expressions of \eqref{xyz} are independent of $h_{rr}$, we have 
\be
T_{rr} = - h_{rr} H \, . 
\ee
The other non-zero components are 
\be
T_{\theta\theta} = (\sin\theta)^{-2} T_{\phi\phi} 
= r^2 \left[2\left({\rm x}H_{\rm x} + {\rm y}H_{\rm y} + {\rm z}H_{\rm z}\right) - H\right] \, . 
\ee
The right-hand side can be simplified by using the fact that 
\be\label{xtor}
4\left({\rm x}H_{\rm x} + {\rm y}H_{\rm y} + {\rm z}H_{\rm z}\right) 
= - r {\rm H}'(r)\, ,    
\ee
where, on the right-hand side, ${\rm H}(r)$ is the function\footnote{We use italic font 
for the Hamiltonian function and roman font for the associated function of $r$.} found from $H({\rm x}, {\rm y}, {\rm z})$ via the $r$-dependence of the variables $({\rm x},{\rm y},{\rm z})$ as given in \eqref{xyz}. We thus find that 
\be 
T_{\theta\theta} = (\sin\theta)^{-2} T_{\phi\phi} = -\frac12 r(r^2{\rm H})'\, . 
\ee
Recalling the definition of $\Theta$ as the trace of the stress-energy tensor, it 
follows from the above results that 
\be\label{defTheta}
\Theta(r) = - \left(r{\rm H}' + 4{\rm H}\right)\, , 
\ee 
which is a formula that we shall use later. 

For convenience, we now set
\be
\NN(r) = e^{\alpha(r)}\, , \qquad h_{rr}(r) = e^{2\beta(r)}\, ,    
\ee
which gives us the following expression for the general   spherically-symmetric spacetime metric: 
\be\label{genmet}
ds^2 = - e^{2\alpha(r)} dt^2 + e^{2\beta(r)} dr^2 + r^2\left(d\theta^2 + \sin^2\theta d\phi^2\right) \, .  
\ee
A computation of the Einstein tensor for this metric shows that the non-zero components match with the non-zero components of the NLED stress-energy tensor, with the result that the Einstein field equations are equivalent to the following component equations:
\be\label{ENLED}
\begin{aligned}
0 &= G_{tt} -(8\pi G)T_{tt} = e^{2\alpha}\left\{ r^{-2}\left[ 1- \left(re^{-2\beta}\right)' \right] -(8\pi G) {\rm H}\right\}
\\
0&= G_{rr} - (8\pi G)T_{rr} = -e^{2\beta} \left\{ r^{-2}\left[ 1- e^{-2(\alpha+\beta)} \left(re^{2\alpha}\right)' \right] -(8\pi G){\rm H}\right\} \\
0&= G_{\theta\theta} - (8\pi G)T_{\theta\theta} = e^{-2\beta} \left[ e^{-2\alpha} \left(r^2 e^{2\alpha}\alpha'\right)' - (r + r^2\alpha')(\alpha+\beta)'\right] \\ 
& \qquad \qquad \qquad \qquad \qquad + (4\pi G) r\left(r^2 {\rm H}\right)' \, . 
\end{aligned}
\ee
The $\phi\phi$-component yields nothing new since
\be
G_{\phi\phi} - (8\pi G) T_{\phi\phi} = \sin^2\theta\left[G_{\theta\theta} - (8\pi G) T_{\theta\theta}\right]\, .
\ee
The $tt$ and $rr$ components are compatible iff
\be\label{albeta}
\alpha+\beta=0\, , 
\ee
and then both equations reduce to the one equation
\be\label{one-eq}
1- \left(r e^{2\alpha}\right)' = (8\pi G) r^2 {\rm H}\, . 
\ee
The remaining independent ($\theta\theta$) equation is now
\be
\left(r^2 e^{2\alpha} \alpha'\right)'  = - (4\pi G)r\left(r^2 {\rm H}\right)'\, , 
\ee 
but this is implied by \eqref{one-eq}, which is therefore the only equation 
that we need solve; we may rewrite it as 
\be\label{one-eq'}
\left(r e^{2\alpha}\right)' = 1- (8\pi G) r^2 {\rm H} \, . 
\ee

Recall that ${\rm H}(r)$ is the function found from $H({\rm x,y,z})$ by using \eqref{xyz}, but this is unchanged by any duality symmetry of $H(D,B)$, which acts linearly on $(q_e,q_m)$. It follows that {\sl any electromagnetic-duality invariance of the NLED Hamiltonian implies a corresponding symmetry in the $(q_e,q_m)$ parameter subspace of  spherically-symmetric black hole spacetime metrics}. In particular, for self-dual theories the metric only depends on $(q_e,q_m)$ through the $U(1)$-duality invariant charge 
\be
Q = \sqrt{q_e^2+ q_m^2}\, . 
\ee
For NLED theories with a $D\leftrightarrow B$ discrete duality symmetry the spacetime metrics for the purely electric and purely magnetic black holes will be the same. 

\subsection{Asymptotically-flat charged black holes}\label{subsec:asymp}

For what follows it is convenient to define a new function $\EE(r)$  by setting 
\be\label{defeta}
re^{2\alpha} = r - 2G\left[M- \EE(r)\right] \, , 
\ee
where the constant $M$ is a free parameter. The key equation of \eqref{one-eq'} now becomes the following first-order ODE for $\EE(r)$:
\be\label{eta1}
\EE'(r) = -4\pi r^2 {\rm H}(r)\, . 
\ee
We may omit the integration constant when integrating this equation because the freedom it represents is precisely the freedom to choose the constant $M$ in \eqref{defeta}.  We thus arrive at the relation 
\be
\label{intEE}
\mathcal{E}(r) = 4\pi \int_r^\infty d\rho \left[\rho^2 {\rm H}(\rho)\right] \, . 
\ee
We are assuming here that $\EE(r)\to 0$ as $r\to\infty$; i.e. that the asymptotic spacetime is not only flat but also empty, which implies that the $r\to\infty$ limit is also a weak-field limit (and, obviously, this requires $H(r)\to 0$ as $r\to \infty$, so the vacuum energy is zero). This limit is either Maxwell or ModMax for a self-dual NLED. For NLED theories that are not self-dual there may be other conformal weak-field limits but we shall ignore this possibility here.

The function $\EE(r)$ has a simple physical interpretation. To see this we observe that \eqref{albeta} implies  
\be 
\NN\sqrt{\det h} = r^2 \sin\theta\, , 
\ee
which is the standard volume density for a 2-sphere of radius $r$ in Euclidean 3-space. Using this in the formula of \eqref{emen} for electromagnetic energy, now specialised to the radial fields of a 
 spherically-symmetric black hole, we find that the total electromagnetic energy {\sl outside} the sphere of radius $r$ is 
precisely $\EE(r)$. The total electromagnetic energy is therefore
\be\label{totalE}
\mathcal{E}_{\rm em} = \lim_{r\to0} \mathcal{E}(r) \, , 
\ee 
but this is finite only if $\mathcal{E}(0)$ is finite; in these cases we can rewrite the 
integral formula of \eqref{intEE} as 
\be\label{EE-finite}
\EE(r) = \EE_{\rm em} - 4\pi\int_0^r d\rho \left[\rho^2 {\rm H}(\rho)\right]\, . 
\ee

It was shown in \cite{Hale:2025ezt} that $\mathcal{E}(r)$ 
is a convex function of $r$ if the NLED stress-energy tensor satisfies the SEC, and this fact was used in much of their analysis. Its significance is that the SEC is implied by the NLED strong-field causality condition \cite{Russo:2024xnh}, and therefore convexity of $\EE(r)$ is a property of all causal NLED theories. The proof of this convexity condition in \cite{Hale:2025ezt} makes use of the well-known fact that the SEC is a condition on the stress-energy tensor that is equivalent, given the Einstein field equations, to $R_{tt}\ge0$. From the Einstein field equations of \eqref{ENLED} we find that 
\be
R_{tt} = \frac{\NN^2 G}{r} \EE''(r)\, , 
\ee
and hence that the SEC is $\EE''(r)\ge0$, which is equivalent to convexity of $\EE(r)$. Of course, this result may be verified directly from the components of the NLED stress-energy tensor in the current black-hole context. Either way, we see that the SEC is satisfied iff
\be\label{SEC}
\EE''(r) \ge 0\, . 
\ee

At this point it is important to appreciate that the SEC is a {\sl weaker} condition than strong-field causality\footnote{The 
original Born theory is an example of a NLED theory for which the SEC is satisfied but the strong-field causality condition is violated \cite{Russo:2024xnh}}. In particular, and in the current context, strong-field causality implies \eqref{SEC} but with the 
additional restriction that equality is allowed only in the NLED vacuum\footnote{More precisely, only for $S=P=0$, but this implies $E=B=0$ for purely radial fields.} \cite{Russo:2024xnh}. For the spherically-symmetric charged spacetimes of relevance here, 
the NLED fields take their vacuum values only asymptotically as $r\to0$ and therefore, for any finite $r$ and (we recall) for zero vacuum energy, 
\be\label{convexE}
\EE''(r) \equiv -4\pi r \left[r{\rm H}'(r) +2{\rm H}(r)\right] \ >0\,  
\ee 
This is a {\sl strict convexity} condition, and we shall see that it is needed for results that depend on strong-field causality.

Another important function for our purposes is the ``effective charge'' function $Q_{\rm eff}(r)$ defined by 
\be\label{EQrel}
2\pi Q^2_{\rm eff}(r) \equiv r\, \mathcal{E}(r)\ .   
\ee
This function is more directly related to the deformation of the RN metric due to NLED interactions since 
\be\label{gtt}
\NN^2 \equiv e^{2\alpha} = 1 - \frac{2GM}{r} + \frac{(4\pi G) Q_{\rm eff}^2(r)}{r^2}\, , 
\ee
with $Q_{\rm eff}=Q$ for Maxwell. It is an important function, for our purposes,  because causality imposes a very simple condition on it, which can be deduced as follows. 

The first-order equation of \eqref{eta1} implies the following second-order equation
\be
\label{secondorder}
\left[Q^2_{\rm eff}(r)\right]'' = -2r^2\left[r{\rm H}' + 4{\rm H}\right]\, ,  \ee
which can be rewritten, using \eqref{defTheta}, in the form
\be
\label{etatheta}
\left[Q^2_{\rm eff}\right]'' = 2r^2 \Theta\, , 
\ee
where (we recall) $\Theta$ is the trace of the stress-energy tensor.
Recalling additionally that the weak-field causality/convexity conditions\footnote{The only other weak-field causality condition is $L_S>0$.} are equivalent to $\Theta\le0$ (for zero vacuum energy and with equality for conformal theories) we deduce that 
\be\label{conc}
\left[Q^2_{\rm eff}(r)\right]'' \le0  \, , 
\ee
where equality holds only for conformal NLED theories. 
This is equivalent to the statement that the function $Q^2_{\rm eff}(r)$ is concave, and strictly concave for any non-conformal NLED theory, in which case it approaches the constant value of the conformal weak-field limit as $r\to\infty$ (this is $Q^2$ if the weak-field limit is Maxwell). It must do so monotonically, and its asymptotic constant value is then also its maximum value. 

If the weak-field limit is ModMax then $Q^2_{\rm eff}(r)$ approaches a constant less than $Q^2$. To see this we observe that the Hamiltonian function $H_{MM}$ of \eqref{ModMax} reduces for $p=0$ (i.e. static fields) to 
\be
H_{MM}= e^{-\gamma}s \qquad (p=0), 
\ee
which yields (for spherically-symmetric  black holes)
\be
r^2{\rm H}_{MM}(r) = e^{-\gamma}\frac{Q^2}{2r^2} \, , 
\ee
and hence
\be
Q^2_{\rm eff}(r) = e^{-\gamma} Q^2 \, \le \, Q^2\, , 
\ee
where the inequality follows from the $\gamma\ge0$ requirement for causality \cite{Bandos:2020jsw}. The only effect of the conformally invariant interactions of ModMax is to reduce the value of the Maxwell charge.

To summarise: we have introduced two functions in this subsection, each satisfying a convexity/concavity causality condition.  These are a strictly convex function $\mathcal{E}(r)$ which equals the electromagnetic energy outside a sphere of radius $r$ (and is therefore asymptotic to zero as $r\to\infty$), and a concave function $Q^2_{\rm eff}(r)$, which is strictly concave for non-conformal NLED and asymptotic to $Q^2$ as $r\to\infty$ if Maxwell is the weak-field limit. These two causality conditions are independent; neither one implies the other. In subsection \eqref{subsec:NSE} we shall see explicitly how convexity of $\EE(r)$ and concavity of $Q^2_{\rm eff}(r)$, equivalently $\Theta(r)\le0$, place different, and complementary, constraints on the behaviour of the black-hole function ${\rm H}(r)$ near $r=0$.

\subsection{Near-origin expansions}\label{subsec:NSE}

As we have seen,  the Hamiltonian function $H$ of any particular NLED theory yields, in the context of spherically-symmetric  black holes, a function ${\rm H}(r)$ that appears on the right-hand side of the Einstein field equations. In a variety of simple examples the leading term of this function as $r\to0$ is a power of $r$:
\be\label{Hrpower} 
{\rm H}(r) \sim r^{-4\nu}\ ,
\ee {\bf [Changed\ to \ Roman\ font]}
for some number $\nu$, with $\nu=1$ for Maxwell. Other behaviors are possible and we discuss this below. 

We can now use causality to restrict the  values of $\nu $.
For example, using \eqref{Hrpower} in \eqref{eta1} we see that 
\be\label{Esim}
\left[\EE(r)\right]' \sim - r^{2(1-2\nu)}  \quad \Rightarrow \quad 
\left[\EE(r)\right]'' \sim (2\nu-1) r^{1-4\nu}\, ,  
\ee
and hence that convexity of $\mathcal{E}(r)$ requires
\be
\nu \ge \frac12\, ,  
\ee 
which is therefore necessary for causality. 

Integrating the expression of  \eqref{Esim} for $\EE'(r)$ we get 
\be\label{firstI}
r\to 0:\quad \mathcal{E}(r) \sim \left\{\begin{array}{cc} 
\EE_{\rm em} - r^{3-4\nu} & \nu <\frac34 \\ -\ln r & \nu=\frac34 \\  r^{-(4\nu-3)} & \nu>\frac34 \end{array} \right. 
\ee
Multiplicative constants have been ignored here, except for their signs, and the integration 
constant $\EE(0)= \EE_{\rm em}$ is included for $\nu<\frac34$ because only in this case is it the leading term; in other words, the total electromagnetic energy $\mathcal{E}_{\rm em} :=\mathcal{E}(0)$ is finite iff 
\be
\nu <\frac34\, . 
\ee
This includes $\nu=\frac12$ and hence Born-Infeld. 

Next, we use \eqref{firstI} to get the following expressions for $Q^2_{\rm eff}(r)$: 
\be\label{Qeff}
r\to 0:\quad Q^2_{\rm eff}(r)  \sim \left\{ \begin{array}{cc} \frac{1}{2\pi}\EE_{\rm em}\, r - r^{4(1-\nu)} & \nu<\frac34 \\
-r\ln r & \nu=\frac34 \\ r^{4(1-\nu)} & \nu>\frac34 \end{array} \right. 
\ee 
Taking two derivatives yields 
\be
r\to 0:\quad \left[Q^2_{\rm eff}(r)\right]'' \sim \left\{\begin{array}{cc} - r^{-2(2\nu-1)} & \nu<\frac34 \\
- r^{-1} & \nu=\frac34 \\ - r^{-2(2\nu-1)} & 1 >\nu> \frac34 \\ 0 & \nu=1 \\ r^{-2(2\nu-1)} & \nu>1 \end{array}
\right.
\ee
From this we see that concavity of $Q^2_{\rm eff}(r)$, and hence causality, requires 
\be
\nu\le1 \, ,  
\label{nuless1}
\ee 
with equality only for conformal theories. The case $\nu=1$ includes Maxwell theory but also special interacting theories with a similar behaviour near $r=0$; an example is given in subsection \ref{subsec:phased}. 

An obvious implication of \eqref{EQrel} is
\be
\EE(0) <\infty \quad \Rightarrow \quad Q^2_{\rm eff}(0) = 0 \, . 
\ee 
The above expressions for these functions near $r=0$ are of course consistent with this implication but they also show that $Q^2_{\rm eff}(0) = 0$ when $\frac34\leq \nu<1 $, for which 
$\EE(0)$ is not finite. Thus, $Q^2_{\rm eff}(0) = 0$ does {\sl not} imply finite $\EE(0)$. 

To summarise: a {\sl necessary} condition for causality is \eqref{nurange-intro}, i.e.
\be\label{nurange}
\frac12 \le \nu \le 1\, ,  
\ee
and $\nu<\frac34$ is required for finite electromagnetic 
energy: $\EE(0)=\EE_{\rm em}$. In subsection \ref{subsec:bh-sd} we shall show that there exists a causal self-dual NLED theory for every value of $\nu$ in the range of \eqref{nurange}.  Both limits of \eqref{nurange} are causality limits but the lower limit 
($\nu\ge \frac12$) is a consequence of the strict convexity of $\EE(r)$ which is itself a consequence of {\sl strong-field} causality, whereas the upper limit $\nu\le1$ is a consequence of {\sl weak-field} causality (since $\nu=1$ for the free-field Maxwell theory and this could become $\nu>1$ as a consequence of arbitrarily-weak interactions). 

\subsubsection{General near-origin behaviour}

The values $\nu= \frac12, \frac34, 1$ of the power-index in our previous discussion based on the power-law behaviour of \eqref{powerlaw} are critical values because they lie at the boundaries or divisions between qualitatively different types of black hole of the kind that we explore later via ``phase diagrams''. What we aim to show now is that these divisions into types can be deduced without having to assume the power-law 
behaviour of \eqref{powerlaw}. 

We consider first how the boundary values $\nu=\frac12,1$ arise
without prior assumptions on the behaviour of the black hole function as $r\to0$. From \eqref{convexE} we see that strict convexity of $\EE(r)$ in the $r\to0$ limit 
is equivalent to 
\be\label{strong}
\left[r^2{\rm H}(r)\right]' < 0 \, . 
\ee
This is the statement that the {\sl least} singular behaviour as $r\to\infty$ of ${\rm H}(r)$ allowed by (strong-field) causality is $H \sim r^{-2}$. In this least-singular case the 
leading term in a small-$r$ expansion of $r^2{\rm H}(r)$ is a non-zero constant and the strict inequality of \eqref{strong} imposes a condition on the sign of the next to leading term.

From \eqref{secondorder} we see that concavity of $Q^2_{\rm eff}(r)$ in the $r\to0$ limit is equivalent to  
\be\label{weak}
\left[r^4{\rm H}(r)\right]' \ge 0\, , 
\ee
where equality is possible only for conformal NLED theories. 
This is the statement that the {\sl most} singular behaviour 
of ${\rm H}(r)$ as $r\to0$ allowed by (weak-field) causality is $H \sim r^{-4}$. In this most-singular case the leading term in the small-$r$ expansion of $r^4{\rm H}(r)$ is a non-zero constant. The inequality of \eqref{weak} is saturated 
for conformal NLED theories, and it constrains the sign of the next-to-leading term if one is present.

We see from the above discussion that the  $\nu=\frac12,1$ cases of the previous analysis based on \eqref{powerlaw} become powers for the least-singular ($\nu=\frac12$) and most-singular ($\nu=1$) behaviour of the black-hole function ${\rm H}(r)$ permitted by causality. Causality thus requires that it increase no slower than $1/r^2$ and no faster than $1/r^4$ as $r\to0$.

Similarly, the division between finite and infinite electromagnetic energy can be defined in terms of the least-singular power for which the electromagnetic energy is infinite. It is straightforward to verify that finite electromagnetic energy requires 
\be
\lim_{r\to 0} \left\{r^3 H(r)\right\}=0\ , 
\ee
The least singular power is therefore $1/r^3$, which corresponds in the analysis based on \eqref{powerlaw} to $\nu= \frac34$. 

We can summarise all the above as follows. Given a black hole function $ {\rm H}(r) $, we define
\begin{equation}
\gamma_n \equiv \lim_{r \to 0} r^{\,n+1} {\rm H}(r)\, ,\qquad n=1,2,3\ .
\end{equation}
Causality imposes the constraints $ \gamma_1 \neq 0 $ and $\gamma_3 \in \mathbb{R}^+\cup \{0\} $. That is, $\gamma_1$
can be finite (but not zero) or can go to infinity at the origin, and $\gamma_3$ must be finite, in particular zero. Finite electromagnetic energy requires $\gamma_2=0$.

We now have the following classification of black-hole ``types'' based on behaviour of ${\rm H}(r)$ near the spacetime singularity:
\begin{itemize}

 \item If $\gamma_3 $ is finite and nonzero, the solution is of Maxwell type, corresponding to $\nu = 1 $.

 \item If $ \gamma_2 = \infty $ and $ \gamma_3 = 0 $, the solution lies in the class $ \tfrac{3}{4} \leq \nu < 1 $,   called ``Intermediate-type I'' in subsection \ref{subsec:phased}.

    \item If $ \gamma_1 = \infty $ and $ \gamma_2 = 0 $, the solution belongs to the class $\tfrac{1}{2} < \nu < \tfrac{3}{4} $, called Intermediate-type II in subsection
    \ref{subsec:phased}.

    \item If $ \gamma_1 $ is finite and nonzero, the solution is of Born-Infeld-type, corresponding to $ \nu = \tfrac{1}{2} $.

\end{itemize}
In subsection \ref{subsec:phased}, we show that this classification applies to black-hole phase diagrams.

As an illustrative example, we consider the self-dual NLED theory defined by the Lagrangian CH-function
\be
\ell(\tau) =T (e^{\tau/T}-1)\, . 
\ee
This satisfies the causality conditions of \eqref{causal-sd}
and therefore defines a causal theory. Using \eqref{relahh},
one finds that the black-hole function has the following 
behaviour as $r\to0$:
\be
{\rm H}(r)\sim -\frac{1}{r^2}\log r\, .  
\ee
For this example we have
\be
\lim_{r\to 0} \left\{r^2 {\rm H}(r)\right\}=\infty\ ,\qquad \lim_{r\to 0} \left\{ r^3 {\rm H}(r)\right\}=0\ .
\ee
This theory, and its black holes, are therefore Intermediate-type II, with $\frac12<\nu<\frac34$.



\subsubsection{Born-Infeld}

Since Born-Infeld is an important special case we give below some details 
of the functions ${\rm H}(r)$, and $\mathcal{E}(r)$ and $Q^2_{\rm eff}(r)$ for this case, including both the large $r$ and small $r$ expansions. We discuss small-$r$ expansions for generic causal NLED with finite $\mathcal{E}_{\rm em}$ in section \ref{sec:global} but some general features are already apparent in the results for 
Born-Infeld. 

We start from the Born-Infeld Hamiltonian function:
\be\label{hyyy}
H_{\rm BI}= \sqrt{(T+2{\rm x})(T+2{\rm y}) -{\rm z}^2} -T \equiv \sqrt{T^2 + 2Ts +p^2} -T\, .   
\ee
Setting $p=0$ leads to 
\be\label{HBI}
{\rm H}_{\rm BI}(r)= T\left\{ \sqrt{1+ \frac{Q^2}{Tr^4}} -1\right\} \, ,  
\ee
and expanding for large $r$ we find that 
\be
r^2{\rm H}_{\rm BI}(r)= \frac{Q^2}{2r^2}\left\{ 1 - \frac{Q^2}{4T r^4} + 
\mathcal{O} \left[ \left(\frac{Q^2}{Tr^4}\right)^2\right]  \right\} \, .  
\ee
This yields the following results: 
\be
\mathcal{E}(r) = \frac{2\pi Q^2}{r} \left\{ 1- \frac{Q^2}{20Tr^4} + 
\mathcal{O}\left[ \left(\frac{Q^2}{Tr^4}\right)^2\right]\right\}\, , 
\ee
and 
\be
(4\pi G)Q^2_{\rm eff}(r) = (4\pi G) Q^2\left\{ 1- \frac{Q^2}{20 Tr^4} + \mathcal{O}\left[ \left(\frac{Q^2}{Tr^4}\right)^2\right]\right\}\, . 
\ee
As expected we have a deformation of the RN metric with 
$Q^2_{\rm eff}(r) <Q^2$. 

Expanding \eqref{HBI} for small $r$ we find that 
\be
{\rm H}_{\rm BI}(r) =-T  + \frac{\sqrt{T}\, Q}{r^2} \left\{ 1 + \mathcal{O}\left(\frac{T\, r^4}{Q^2}\right) \right\}\, .   
\ee 
Using this result and the fact that for finite $\EE_{\rm em}$ (as is the case for BI) we have 
\be\label{E-H}
\EE(r) = \EE_{\rm em} - 4\pi\int_0^r r^2 {\rm H}(r) dr\, , 
\ee
we find that 
\be
\label{EEBBI}
\mathcal{E}(r) = \mathcal{E}_{\rm em} +\frac{4\pi}{3}T r^3 - 4\pi \sqrt{T}\, Q\,  r \left\{ 1
+ \mathcal{O} \left(\frac{T\, r^4}{Q^2}\right) \right\} \, , 
\ee 
where (for this BI case) \cite{deOliveira:1994in}
\be\label{BIem}
\mathcal{E}_{\rm em} =\frac{2}{3}\,  \Gamma\Big(\frac14\Big)^2\sqrt{\pi}\,  T^\frac14 Q^\frac32 .
\ee
The second term in \eqref{EEBBI} is the energy in the ball of radius $r$ for energy density $T$, and its origin is the constant $-T$ term in $H$ that has its sign reversed in \eqref{E-H}. 
The higher corrections to the term linear in $r$ come from inverse powers of $s$ in the expansion of the Hamiltonian at large $s$.  Using \eqref{EEBBI} in \eqref{EQrel} one finds the following small-$r$ expansion of the effective charge function:
\be
(4\pi G) Q^2_{\rm eff}(r) = \left(2G\mathcal{E}_{\rm em}\right) r+\frac{8\pi}{3} \, GT r^4 
- (8\pi G)\sqrt{T}\, Q\,  r^2 \left\{ 1 + \mathcal{O}\left( \frac{T\, r^4}{Q^2}\right)  \right\} \, . 
\ee 
Notice that  the $\mathcal{O}(Tr^4/Q^2)$ terms in these expansions can be neglected when $T r^4 \ll Q^2$, which is equivalent to 
\be
 r\ll \sqrt{[\sqrt{G} Q]\,  \ell_{gB}}\, . 
 \ee
 In other words, $r$ must be much less than the geometric mean of the two `gravitational' length scales $\sqrt{G}Q$ and $\ell_{gB}$  associated to $Q$ and $T$ respectively; on dimensional grounds, this result for Born-Infeld applies to any non-conformal NLED theory.   Recalling, from the Introduction, that qualitatively significant deviations from RN geometry should be expected when $\sqrt{G}Q$  is approximately  equal to (or less than) $\ell_{gB}$, we conclude that the validity of small-$r$ expansions in any investigation of  these deviations  requires $r\ll \ell_{gB}$. 


\subsection{Electric field}\label{subsec:electric}

Born's original NLED theory has the feature that it puts an upper bound on the strength $|E|$ of the electric field; specifically $|E|^2 \le T$. This is also true of the Born-Infeld theory, although the maximum value is larger for non-zero magnetic field. This property is potentially of importance to charged black holes since the electric field strength blows up at the central singularity of the RN black-hole solution to the Einstein-Maxwell equations, whereas this does not occur if Maxwell electrodynamics is replaced by Born-Infeld. We now aim to investigate the behaviour of the electric field near the central singularity of an electrically charged black hole for a generic causal NLED theory. 

For zero magnetic field, the expression of \eqref{DD} for $D$ as the partial derivative of $\LL(S,P)$ with respect to the electric field $E$ reduces to 
\be\label{DLE}
D^i = \frac{\sqrt{\det h}}{\NN}h^{ij} L_S E_j\, ,  
\ee
where now $S= \frac12 \NN^{-2}|E|^2$ and $P=0$. For a static and spherically-symmetric electric field configuration in the general black hole spacetime metric of \eqref{genmet}, we find that 
\be
\frac{D^r}{r^2\sin\theta} = L_SE_r \, , \qquad  S= \frac12 E_r^2\, . 
\ee
Since spherical symmetry requires $D^r= q_e \sin\theta$, we have
\be\label{Err}
E_r  = \frac{q_e}{r^2 L_S} \, ,  
\ee
and taking a derivative with respect to $r$ on both sides one finds that 
\be
\label{deree}
\left[L_S + 2S L_{SS}\right]\frac{dE_r}{dr} = - \frac{2q_e}{r^3}\, .  
\ee 
We may assume here, without loss of generality, that $q_e>0$, in which case $E_r\ge 0$. 

Two necessary conditions for causality of any NLED theory are \cite{Schellstede:2016zue}
\be\label{causal-L}
L_S> 0 \, , \qquad L_{SS} \ge 0 \, .   
\ee
However, since $L_S\ge1 $ in the conformal weak-field limit (with equality when this is Maxwell), $L_S$ must satisfy the slightly stronger causality conditions\footnote{Recall that these have been derived for $B=0$. In fact, they are the causality conditions of \eqref{causal-sd} for self-dual NLED generalized (as necessary causality conditions) to generic NLED.} 
\be\label{cfsd}
L_S \ge1\, , \qquad L_{SS}\ge 0 \, , 
\ee 
with equality (for a Maxwell weak-field limit) as $S\to 0$ (i.e. $r\to0$ in the current black-hole context). Both pairs of conditions imply that $L_S$ {\sl is a positive non-decreasing function of $S$}, but its minimal value at $S=0$ is more strongly constrained by \eqref{cfsd}. 

\begin{figure}[h!]
 \centering
\includegraphics[width=0.6\textwidth]{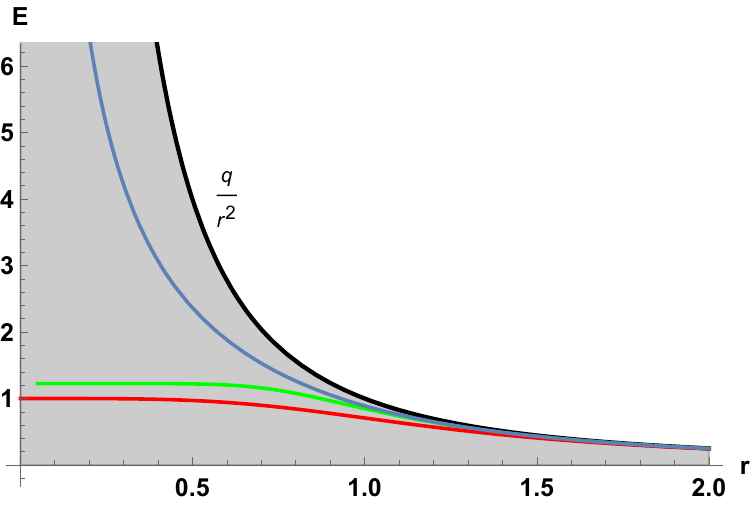}
 \caption
 {The electric field $E_r(r)$ for Maxwell (black),  Born-Infeld (red) and two other examples of causal theories with $E_r\sim 1/r$ (blue)  and $E_r\to E_0$ (green). For any causal NLED, $E_r(r)$ is a monotonically decreasing function of $r$ that is less than $Q/r^2$ for all finite $r$, and 
 hence lies entirely in the grey-shaded area.}
 \label{elecfig}
 \end{figure}

The implications for the electric field are illustrated in Fig. \ref{elecfig}. For example,  $E_r$ is a monotonically decreasing function of $r$ in any causal NLED; this follows immediately from \eqref{deree} since causality requires  $(L_S+ 2SL_{SS})>0$ when $B=0$. In fact, $(L_S +2S L_{SS})>1$ for positive $S$, and this has the further implication that the graph of the function $E_r(r)$ for any non-conformal causal NLED lies below this graph for Maxwell (or ModMax if that is the weak-field limit) except asymptotically as $r\to\infty$ (where they meet). This implies that the rate of increase of $E_r$ as $r\to 0$ is less than it is for Maxwell; in particular, it can happen that $E_r$ remains finite as $r\to0$. 

A Hamiltonian version of the formula \eqref{Err} for the electric field can be found 
by taking \eqref{defE}, with $B=0$, as the starting point. Instead of \eqref{DLE} we now have
\be\label{EHD}
E_i = \left(\frac{\NN}{\sqrt{\det h}}\right) h_{ij} H_{\rm x} D^j\,  , 
\ee
which leads (since $H_{\rm x} = H_s$ for ${\rm y}={\rm z}=0$) to
\be
E_r = H_s \frac{q_e}{r^2} \, , 
\ee 
where now $s=\frac12 (D^r)^2$. A comparison of this result with \eqref{Err} shows that (cf. \eqref{CHderivs})
\be
\label{LHH}
L_S H_s=1\ , 
\ee
which is essentially the relation of \eqref{CHderivs} (expressed there in terms of derivatives of the CH functions of self-dual NLED theories).  

As for self-dual NLED in the previous subsection, if 
$H(s) \sim s^\nu$ then 
$H_s\sim s^{\nu-1}$ and we can convert this into a function of $r$ using \eqref{stor}.
This leads to the following behaviour of $E_r$ near the central singularity:
\be\label{Enz}
E_r(r) \sim \frac{1}{r^{2(2\nu-1)}}\, . 
\ee
Given the $\nu\ge\frac12$ restriction required by causality, the electric field will 
remain finite as $r\to0$ iff $\nu=\frac12$, even though (as we saw above) the total electromagnetic energy remains finite provided $\nu<\frac34$. The Born-Infeld theory illustrates this since, in this special case, 
\be
E_r = \frac{q_e \sqrt{T}}{\sqrt{Tr^4 + q_e^2}} \quad \rightarrow\quad  
\sqrt{T}\, . 
\ee
In contrast, for $\frac12 <\nu <\frac34$ (which is the remainder of the range of the exponent $\nu$ for which the electromagnetic energy is finite) $E_r(r)$ blows up as $r\to 0$; this is a reminder that finite energy does {\sl not} require a finite electric field at $r=0$. Finally, for $\frac34 \le \nu<1$ both $E_r(r)$ and $\mathcal{E}(r)$ blow up as $r\to0$ (as also occurs in the $\nu=1$ limit, which includes the  Maxwell case).

\subsection{Black holes for self-dual NLED} \label{subsec:bh-sd}

For self-dual NLED it is simple to find the black-hole function ${\rm H}(r)$ from the Hamiltonian function $H$ because the latter must be a function of $(s,p)$ only, and $p=0$ for static field configurations. Thus, $H$ can be viewed as a function of $s$ only, which we call $H(s)$, and then ${\rm H}(r)$ is found by the substitution 
\be\label{stor}
s= \frac{Q^2}{2r^4}\ .
\ee 
We see from this that the $s\to\infty$ limit of $H(s)$ corresponds to the $r\to0$ limit of ${\rm H}(r)$. If we now suppose that 
\be\label{Hbehave}
H(s) \sim  s^\nu  \qquad (s\to\infty) \, , 
\ee
then we find the power-law behaviour of ${\rm H}(r)$ postulated in \eqref{Hrpower}.  

We summarised in subsection \ref{subsec:sd} the construction of both the Lagrangian and Hamiltonian functions of Lorentz-invariant self-dual NLED theories in terms of one-variable 
CH-functions, and we explained how the causality constraints on generic NLED theories then reduce to much simpler constraints on these functions. In particular, the Hamiltonian function $H(s,p)$ is expressed in terms of the Hamiltonian CH-function $\mathfrak{h}$ such that  
\be
\mathfrak{h}(s) = H(s,0) \equiv H(s)\, .  
\ee
This allows us to rewrite the causality constraints of \eqref{hfrak-c} as
\be\label{causalityHs}
0<H_s \leq 1 \, , \qquad H_{ss} \leq 0\, ,\qquad  H_s+2sH_{ss} >0\ .
\ee
Assuming the power-law behaviour of \eqref{Hbehave} we find that the first two inequalities are satisfied if $0< \nu\le 1$. The third inequality (now allowing for its saturation because we are applying it in the $s\to\infty$ limit) requires $\nu\ge\frac12$. We thus recover the causality restriction of \eqref{nurange} from the large $s$ behaviour of $H$. 

Now consider the following choice of Hamiltonian CH-function \cite{Russo:2024ptw}:
\be\label{frakh1}
\mathfrak{h}(\sigma)= T \left\{ \left(1+ \frac{\sigma}{\nu T} \right)^\nu -1\right\} = \sigma + \mathcal{O}(\sigma^2)\, .   
\ee
Notice that the $\nu=1$ case is Maxwell and the $\nu=\frac12$ case is Born-Infeld. This CH-function yields, via \eqref{HamCH}, the Hamiltonian function $H(s,p)$ of a Lorentz-invariant self-dual NLED theory, with  
\be
H(s) \equiv \mathfrak{h}(s) \ \sim \ s^\nu \qquad (s\to\infty). 
\ee
A necessary condition for causality of the NLED defined by $H(s,p)$ is therefore 
the constraint \eqref{nurange} on the range of $\nu$. Moreover, for this particular self-dual NLED theory the constraint on the range of $\nu$ is also {\sl sufficient} for causality. This is important because it shows that {\sl the power-law behaviour that we assumed for the $r\to0$ limit
of the``black-hole'' function ${\rm H}(r)$ is actually realised in a simple causal model for all values of $\nu$ allowed by causality}. 

Consider now the large-$\sigma$ expansion of the function $\frak{h}(\sigma)$ defined by \eqref{frakh1}:
\be\label{frakh2}
\frak{h}(\sigma) = T\left\{\left(\frac{\sigma}{\nu T}\right)^\nu \left[ 1 + \nu \left(\frac{\sigma}{\nu T}\right)^{-1} + \dots\right] -1\right\}\, . 
\ee
Using the fact that $\frak{h}(\sigma)$ and $H(s)$ are identical functions of their respective independent variables, and the relation \eqref{stor}, we find the following 
small-$r$ expansion of ${\rm H}(r)$:
\be\label{frakh3}
r^2 {\rm H}(r) = -Tr^2 + T\left(\frac{Q^2}{2\nu T}\right)^\nu \frac{1}{r^{2(2\nu-1)}} + 
T\nu \left(\frac{Q^2}{2\nu T}\right)^{\nu-1} r^{2(3-2\nu)} + \ \dots 
\ee
The first term comes from the subtraction of $T$ in \eqref{frakh2} needed for zero vacuum energy. The 
second term is a constant for $\nu=\frac12$; it is singular as $r\to0$ for $\nu>\frac12$ but this singularity is integrable for $\nu<\frac34$.  The third term (and all higher-order terms) tends to zero faster than $r^2$ as $r\to 0$ because $\nu<1$. Since the electromagnetic energy density is finite for $\nu<\frac34$, we may use the above small-$r$ expansion in \eqref{EE-finite} to arrive at the following small-$r$ expansion of $\EE(r)$:
\be\label{EEsmall-r} 
\EE(r)=\EE_{\rm em}-c_\nu  Q^{2\nu}T^{1-\nu}\, r^{3-4\nu}+\frac{4\pi}{3}T r^3 + \mathcal{O}\left( r^{7-4\nu}\right)\, . 
\ee
The term proportional to $r^3$ has the same origin as the similar term in \eqref{EEBBI}.
The electromagnetic energy can be found exactly as can the coefficient $c_\nu$:
\be\label{Enu}
\EE_{\rm em}=\frac{2^{\frac{5}{4}}  \Gamma
   \left(\frac{1}{4}\right)  \sin (\pi  \nu ) \Gamma(\nu+1)\Gamma
   \left(\frac{3}{4}-\nu \right)}{3\, \nu ^{\frac34}}\ Q^{\frac32} T^{\frac14}\ ,\quad c_\nu \equiv \frac{  2^{2-\nu } \pi\nu ^{-\nu } }{3-4 \nu }\ . 
\ee
We see that the electromagnetic energy is indeed finite for $\nu<\frac34$. 
Notice that $c_\nu$ is positive because $\nu<\frac34$; for the same reason the term with coefficient $c_\nu$ is zero at $r=0$. 

The small-$r$ expansion in  \eqref{EEsmall-r} and \eqref{Enu} agrees 
for $\nu=\frac12$ with \eqref{EEBBI} and \eqref{BIem} for Born-Infeld, as it must since 
this {\sl is} the Born-Infeld case. This expansion was proposed in \cite{Hale:2025ezt}, to linear order in $r$, as generic for cases with finite electromagnetic energy, but we can now see that this linear term is particular to the 
$\nu=\frac12$ (BI-type) cases, and is replaced by a lower fractional power for all 
other cases.

\subsubsection{A Lagrangian perspective}

We have now seen how the behaviour of ${\rm H}(r)$ as $r\to0$ is related to the behaviour of $H(s)$ as $s\to\infty$ or, equivalently, to the behaviour of the Hamiltonian CH-function $\mathfrak{h}(\sigma)$ as $\sigma\to\infty$, but how is it related to the Lagrangian CH-function $\ell(\tau)$? 
We summarised in subsection \ref{subsec:sd} the relation between the Hamiltonian 
and Lagrangian CH-functions. Using the relations \eqref{relah} and/or \eqref{relahh}, 
we find that, as $\sigma\to \infty$, 
\be
\ell(\tau) \sim (2\nu-1)\sigma^\nu \, , \qquad \tau \sim \sigma ^{2\nu-1} \, , 
\ee
We see again that the $\nu=\frac12$ case is special, so we postpone discussion of it.

For $\nu>\frac12$ we have 
\be\label{ellr0} 
\sigma\to\infty \ \Rightarrow \  \tau\to \infty \ \Rightarrow  \ \ell(\tau) \sim  \tau^\beta \, , 
\ee
where $\nu$ and $\beta$ related by $2\beta \nu = \beta+\nu$. For example
\be\label{beta}
\beta = \nu/(2\nu-1) = \left\{\begin{array}{cc} \beta =1 & \nu=1 \\ \beta= \frac32 & \nu= \frac34\\
\beta \to\infty & \nu\to\frac12 \end{array} \right. 
\ee 

From this we see that the electric field diverges as $r\to0$ for all values of $\beta$, but the 
electromagnetic energy is finite for $\beta>\frac32$.  
Examples of Lagrangian CH-functions defining causal self-dual NLED theories are
\be\label{merem}
\ell(\tau) =
T\left\{ \left[1+\frac{\tau}{\beta T}\right]^\beta-1\right\}\ , \qquad \beta \ge 1\ .
\ee
Black hole solutions, and their thermodynamic properties,  for self-dual NLED theories with CH-functions of this form, and of the form \eqref{arem} given below, were recently discussed in 
\cite{Babaei-Aghbolagh:2025tim}.

The $\nu=\frac12$ cases, for which the electric field is finite at $r=0$, are very different because there is now some maximum value of $\tau$  ($\tau_{\rm max} >0$) such that 
\be
\sigma\to\infty \quad \Rightarrow \quad  \tau\to\tau_{\rm max}\ :\quad 
\ell(\tau)\to \ell(\tau_{\rm max})\ .
\ee
A simple one-parameter family of causal NLED theories of this type is found from the
Lagrangian CH-function\footnote{This has been used in \cite{Babaei-Aghbolagh:2025tim} for a study of Einstein-NLED black holes; the general solution can be expressed in terms of a hypergeometric function.}
\be\label{arem}
\ell(\tau) =T\left\{ 1-\left[1-\frac{\tau}{\alpha T}\right]^\alpha\right\}\ , 
\qquad 0<\alpha < 1\, . 
\ee
Notice that $\tau_{\rm max} = \alpha T$, and $\ell(\tau_{\rm max})= T$; the restriction of the range of the parameter $\alpha$ is required by causality\footnote{Except for $\alpha =1$, which we exclude because it yields the Maxwell theory, for which $\nu=1$.}. The $\alpha=\tfrac12$ case yields the Born-Infeld theory and hence\footnote{For self-dual NLED the index $\nu$ depends only the choice of NLED theory and is independent of the electric to magnetic charge ratio.} $\nu=\frac12$; in fact, 
$\nu=\frac12$ for all values of $\alpha$ is the allowed range. This can be verified as follows.

From the expansion of $\ell(\tau)$ about $\tau_{\rm max}$, and the relations between the CH functions 
$\ell(\tau)$ and $\frak{h}(\sigma)$ (and their independent variables) summarised in subsection \ref{subsec:sd}, we find the following large-$\sigma$ expansion of $\frak{h}(\sigma)$:
\be\label{h-alpha}
\begin{aligned} 
\frak{h}(\sigma) &= 2 \sqrt{\alpha T\sigma} \left\{ 1 - \frac{1}{2\alpha}\left(\frac{\sigma}{\alpha T}\right)^{-\frac12} + \left(\frac{1-\alpha}{2\alpha}\right) 
\left(\frac{\sigma}{\alpha T}\right)^{-\frac{1}{2(1-\alpha)}} + \ \dots\right\} \\
& \sim (T\sigma)^\frac12 \qquad (\sigma\to\infty), 
\end{aligned}
\ee
and hence $H(s) \sim s^\frac12$ as $s\to\infty$, which implies $\nu=\frac12$. 
We also get this way the following small-$r$ expansion of the black-hole function:
\be\label{r2H-nu}
r^2{\rm H}(r)= -Tr^2 + \sqrt{2\alpha T}\, Q +   (1-\alpha)T \left(\frac{Q^2}{2\alpha T}\right)^{-\frac{\alpha}{2(1-\alpha)}} \, r^{\frac{2}{(1-\alpha)}} + \ \dots 
\ee
Using this result in the procedure leading to \eqref{EEBBI} for Born-Infeld 
we find that 
\be\label{aremE}
\EE(r) = \EE_{\rm em} + \frac{4\pi T}{3} r^3 - 4\pi\!  \sqrt{2\alpha T}\, Q r 
\left\{1 + \frac{(1-\alpha)^2}{2\alpha (3-\alpha)} 
\left(\frac{2\alpha T}{Q^2}\right)^{\frac{1}{2(1-\alpha)}} r^{\frac{2}{1-\alpha}} +  \dots \right\},
\ee
with
\be
\EE_{\rm em}=\frac{ \pi ^2 Q^{\frac32}
   \Gamma
   \left(\frac{3}{2}-\frac{\alpha }{2}\right)
   (8\alpha  T)^{\frac14}}{\Gamma
   \left(\frac{7}{4}\right)
   \Gamma
   \left(\frac{7}{4}-\frac{\alpha }{2}\right)}
  \label{energyalfa}
   \ee
which is consistent, for $\alpha=\frac12$, with the result of \eqref{EEBBI} for Born-Infeld.

\subsubsection{Causality and the SEC revisited}

We explained in subsection \ref{subsec:asymp} that 
strong-field causality implies a `strict' version 
of the SEC for a NLED stress-energy tensor. It is 
simple to verify this for self-dual NLED by using 
the CH-function formulation reviewed in 
subsection \ref{subsec:sd}, as we now explain.  

We return to the SEC condition of \eqref{SEC}, which we may rewrite as
\be\label{Strict}
r{\rm H}'(r) + 2{\rm H}(r) \le 0\, . 
\ee
Using  the relation \eqref{stor}, we have 
\be
r{\rm H}'(r) = -4 sH_s(s)\, ,  
\ee
which allows us to rewrite \eqref{Strict} as
\be
2s H_s(s) -H(s) \ge 0\, . 
\ee
Using \eqref{Hsss} we see that the Hamiltonian CH-function 
$\frak{h}(\sigma)$ must satisfy the same inequality, but also 
from \eqref{relah} that this is equivalent to the following very simple condition 
on the Lagrangian CH-function $\ell(\tau)$:
\be\label{SECL}
\ell(\tau) \ge 0 \, ,  
\ee
which was identified in \cite{Russo:2024xnh} as equivalent to the SEC for self-dual NLED. 

In the black hole context $\tau\to0$ corresponds to $r\to\infty$. This is because $\tau = sH_s^2(s) \sim s$ as $s\to0$, and therefore $\tau\to0$ when $s\to0$. We therefore expect equality in \eqref{SECL} at $\tau=0$, and this is the condition for zero vacuum energy, $\ell(0)=0$. We may now ask whether equality in \eqref{SECL} is possible for some $\tau>0$. This is equivalent to asking whether the inequality of \eqref{SEC} can be saturated other than by taking the $r\to\infty$ limit. 
The answer is no for a causal theory because the causality conditions of \eqref{causal-sd} imply $\ell(\tau)>0$ for $\tau>0$, which accords with our conclusion in subsection \ref{subsec:asymp} that causality implies the `strict' version of the SEC.

\subsection{An explicit causal non-self-dual case}\label{subsec:nonsd}

While most of our results apply to generic causal NLED theories, most of the explicitly known causal NLED theories are self-dual, and these fail to illustrate some novel features that can arise for generic non-self-dual theories. To compensate for this, we now consider 
the following one-parameter family of Lagrangian functions introduced in \cite{deOliveira:1994in}:
\be\label{Lq}
L^{(q)} = \frac{T}{2q}\left\{1- \left(1-\frac{2S}{T} -\frac{P^2}{T^2}\right)^{\!q}\right\} \, . 
\ee
When the causality constraints found in 
\cite{Schellstede:2016zue} are applied to this NLED family one finds only the following constraint on the range of the parameter $q$ \cite{Russo:2024kto}:
\be
\frac12 \le q <1 \, . 
\ee
The $q=\frac12$ case is Born-Infeld, and this is the only self-dual 
NLED of the family \cite{Russo:2024kto}. 
The weak-field expansion to first non-linear order in $(S,P)$ is 
\be
L^{(q)} = S+ \frac{1}{2T}\left[ 2(1-q)S^2  + P^2\right]  + \mathcal{O}(1/T^2)\, , 
\ee
and the corresponding Hamiltonian function is 
\be
H^{(q)} = s - \frac{1}{2T} \left[2(1-q)s^2 -p^2 + 4(2q-1){\rm xy}\right] + \mathcal{O}(1/T^2).
\ee
As expected, this is $U(1)$-duality invariant only for $q=\frac12$. Although a $Z_2$-duality invariance for any $q$ survives to this order in the expansion, it does not survive in the next order. 

We shall focus on the $q= \frac34$ case, for which the full Hamiltonian function can be found explicitly:
\be\label{Hq34}
H^{(\frac34)}({\rm x},{\rm y},{\rm z}) = \frac{2T}{3} \left\{ \frac{\left[\Sigma + 2(p^2 +2T{\rm x})\right] 
\sqrt{\Sigma - (p^2+2T{\rm x})}}{\left[2T(T+2{\rm y})\right]^\frac32} -1\right\}\, , 
\ee 
where 
\be
\Sigma = \sqrt{(p^2 + 2T{\rm x})^2 + 4T(T+2{\rm y})^3}\, , 
\ee 
and (we recall) $p^2 =4{\rm xy}-{\rm z}^2$. 
For static configurations we have $p=0$, and then 
\be
\label{pzerop}
H^{(\frac34)}({\rm x},{\rm y},\sqrt{4{\rm xy}}) =\frac{2 T}{3}\left\{\frac{(K+2{\rm x})\sqrt{K- {\rm x}}}{
   (T+2{\rm y})^\frac32}-1\right\}\, , 
\ee
where 
\be
K := \sqrt{ {\rm x}^2+T^2\left(1+\frac{2{\rm y}}{T}\right)^3}\, . 
\ee
For the particular case of ${\rm y}=0$ (and hence ${\rm x}=s$) we have
\be
\begin{aligned}
H_e(s) :=& \ H^{(\frac34)}(s,0,0) \\
=& \ \frac{2T}{3} \left\{\left[\sqrt{1+ s^2/T^2} +2s/T\right] \sqrt{\sqrt{1+ s^2/T^2}- s/T} \ -1\right\} \\
\sim & \ \sqrt{2T} s^\frac12 \qquad (s\to\infty).
\end{aligned}
\ee
From the large-$s$ behaviour we can read-off the power-index $\nu$ that determines the $r\to0$ behaviour of the black-hole function ${\rm H}(r)$: it is $\nu=\frac12$, and we thus conclude that {\sl electric} black holes are of BI-type. For {\sl magnetic} black holes we set ${\rm x}=0$ in \eqref{pzerop} (and hence ${\rm y}=s$) to get
\be\label{magnetic}
\begin{aligned}
H_m(s) := & \ H^{(\frac34)}(0,s,0)=\frac{2T}{3}\left\{(1+2s/T)^\frac34-1\right\} \\
\sim & \ \frac{2^\frac74}{3} T^\frac14 s^\frac34 \qquad (s\to\infty). 
\end{aligned}
\ee
Now we have $\nu=\frac34$, which implies an infinite energy in the magnetic field. The corresponding black-hole functions near $r=0$ are 
\be\label{Hem}
{\rm H}_e(r) = \frac{T^\frac12 q_e}{r^2} + \ \mathcal{O}(r^0)\, , \qquad 
{\rm H}_{m}(r) =  \frac{2T^\frac14 q_m^\frac32}{3r^3} + \ \mathcal{O}(r^0) \, . 
\ee

Let us now consider a dyonic black hole; i.e. $q_eq_m >0$ since we may assume that both $q_e$ and $q_m$ are positive. Using the expressions of \eqref{xyz} for ${\rm x}$ and ${\rm y}$ in 
\eqref{pzerop} one finds the following small-$r$ expansion of the dyonic black-hole function:
\be\label{dyonH}
{\rm H}_d(r) =\frac{2 T^{\frac14}\ q_m^{3/2}}{3 r^3}+\frac{T^{\frac34} q_e^2/q_m^\frac32}{2 r} + 
\ \mathcal{O}(r^0)\, . 
\ee
This reduces to the small-$r$ expansion of ${\rm H}_m$ when $q_e=0$, but the electric charge
contribution is very different from ${\rm H}_e$. This is not a contradiction because we cannot set $q_m=0$ in \eqref{dyonH}; the expansion in powers of $1/r$ becomes unreliable, at any $r$, as $q_m\to0$. This indicates a restructuring of the spacetime metric near $r=0$ whenever the magnetic charge goes to zero. 

We can exchange the roles of electric and magnetic charge by performing the interchange ${\rm x} \leftrightarrow {\rm y}$ in the Hamiltonian function of \eqref{Hq34}. This defines the $Z_2$ magnetic dual of the NLED theory defined by the Lagrangian function $L^{(\frac34)}$, which we would expect to be a causal theory for which there is a similar discontinuity of the spacetime structure of dyonic black holes in the limit of zero electric charge.

\section{Causality implies spacetime singularity}\label{CimpS}

For NLED theories defined by a Lagrangian function $L(S)$, it was shown by Bronnikov many years ago that any regular (singularity-free) spherically-symmetric charged black-hole solution of the Einstein-NLED equations can have only magnetic charge \cite{Bronnikov:2000vy}. For any 
 spherically-symmetric black-hole solution of the Einstein-NLED equations the scalar $S$ becomes a function of $r$ only, and $L(S)$ therefore becomes a function of $r$, which we write as ${\rm L}(r)$. It was also shown in \cite{Bronnikov:2000vy} that a necessary condition for the existence of a regular magnetically-charged black-hole solution is finiteness of ${\rm L}(r)$ as $r\to0$, which is equivalent to finiteness of $L(S)$ as $|S|\to\infty$.

As we now know, all Born-type NLED theories, i.e. those defined by  a Lagrangian function $L(S)$ (excepting Maxwell) are acausal \cite{Schellstede:2016zue}. The possibility of regular charged black holes for specific examples within the larger  ``Plebanski class''  of NLED theories with  Lagrangian functions $L(S,P)$  was investigated relatively recently by Bokulic et al. \cite{Bokulic:2022cyk} but only additional constraints on $L$ were found.  Around the same time,  Bronnikov 
extended to this larger class of NLED theories his earlier result that any {\sl regular}  spherically-symmetric charged black hole must have magnetic charge only  \cite{Bronnikov:2022ofk}; this  immediately excludes the possibility of regular  black-hole solutions for any self-dual NLED theory. 

For purely magnetic 
black holes we have 
\be
S= - \frac{q_m^2}{2 r^4}\, , \qquad P=0\, ,  
\ee
and $L(S) := L(S,0)$ reduces to a function ${\rm L}(r)$ which is the same as it would be for the simpler (and acausal) NLED theory 
defined by $L(S)$.  In \cite{Bronnikov:2022ofk} Bronnikov  also extends to generic NLED theories\footnote{Actually, the Lagrangian functions considered were $L(S,J)$ where 
$J=4(2S^2 + P^2)$. This `new' class of NLED theories is the parity invariant subset of the class defined by $L(S,P)$, and in the current context ($P=0$) this restriction is unimportant.} his earlier finiteness condition on ${\rm L}(r)$ as $r\to 0$, which can again be expressed as 
\be\label{B2}
\lim_{|S|\to\infty} |L(S)|  < \infty \,  , 
\ee 
but now for $L(S):= L(S,0)$ with $S<0$. As we now show, this condition cannot be satisfied by any causal NLED for which $L(S,P)$ has a weak-field expansion. 
The proof is similar to  one presented in \cite{Bronnikov:2022ofk} for Lagrangian functions $L=L(S)$, but now valid for $L(S) := L(S,0)$. For simplicity, we assume in what follows that Maxwell is the weak-field limit. 

Although the full causality conditions on $L(S,P)$ must be satisfied by any causal NLED with a weak-field expansion, it suffices for our purposes to show that any one of these causality conditions is violated if \eqref{B2} is true.
We shall focus on the convexity conditions $L_S>0$ and $L_{SS}\ge0$, where equality holds only for conformal NLED theories \cite{Schellstede:2016zue}. Additionally, since we assume a Maxwell weak-field limit such that $L_S(0) =1$ in our conventions (which include unit speed of light) we have for all non-conformal 
NLED theories the following two necessary conditions for causality:
\be\label{Bc}
L_S \ge 1 \, , \qquad L_{SS} >0 \, , 
\ee
where $L_S=1$ only in the $S\to0$ limit. Since $S<0$ for zero electric field, 
these inequalities state that $L(S)$ decreases as $|S|$ increases, and that the rate of decrease is positive. Thus, provided only that $L(0)$ is finite (which allows for the possibility of a non-zero vacuum energy) $L(S)$ must be negative for sufficiently large $|S|$ and then $|L(S)|$ must increase at an increasing rate as $|S|$ increases. This implies that $|L(S)| \to\infty$ as $S\to -\infty$, and hence that the condition \eqref{B2} (required for the existence of a purely magnetic spherically-symmetric charged black hole solution of the Einstein-NLED equations) cannot be satisfied. Because we did not assume zero vacuum energy, this result applies irrespective of the value of the cosmological constant. The irrelevance of a cosmological constant for this issue is also explained briefly in \cite{Bronnikov:2022ofk}.

Taken together with the no-go result of Bronnikov for regular black holes with non-zero electric charge, and his (updated) finiteness condition on $L$ for purely magnetic black holes, the above result settles the issue of the existence of regular charged black holes for any causal NLED theory with a 
weak-field expansion: they do not exist.

\subsection{A direct proof}

Using the solution \eqref{defeta}, \eqref{intEE}, the curvature invariants $R$ and $R_{\mu\nu}R^{\mu\nu}$ 
can be expressed in terms of ${\rm H}(r)$. One finds
\bea
R &=& -8\pi G\,  \Theta(r) \ =\  8\pi G \left(r{\rm H}'+ 4{\rm H}\right)\ ,
\\
R_{\mu\nu}R^{\mu\nu}&=& 32\pi^2 G^2\left[ \left(r{\rm H}'+ 2{\rm H}\right)^2+4{\rm H}^2\right]\ .
\eea
Obviously, these two curvature scalars vanish if ${\rm H}(r)\equiv 0$. We then have a (zero-charge) Schwarzschild black-hole spacetime and the singularity appears in (Riemann)$^2$. 

For a {\sl charged} black hole ($Q>0$) we have ${\rm H}(r)>0$.  Using \eqref{defTheta} and \eqref{convexE}, we may rewrite the above equations as follows:
\bea
R &=& -8\pi G\,  \Theta \ge0 \quad  \\
R_{\mu\nu}R^{\mu\nu}&=& \frac{32\pi G^2}{r} \left( \pi r\, \Theta^2 + {\rm H}\, \EE''\right) \ >0
\eea
where the inequalities follow from the fact that $\Theta\le0$
is required by weak-field causality and $\EE''>0$ is required for strict convexity of $\EE$, which is a strong-field causality condition.

The scalar curvature $R$ is zero for conformal theories such as Maxwell and ModMax because in these cases $\Theta\equiv0$, but $R_{\mu\nu}R^{\mu\nu}$ is still non-zero, and positive 
for any charged black hole.
For any non-conformal causal NLED theory both $R$ and 
$R_{\mu\nu}R^{\mu\nu}$ are strictly positive, and we need to investigate how they behave near $r=0$.

For ${\rm H}(r)\sim r^{-4\nu}$, the behaviour near the origin is
\bea
R &\approx & 4G\, r^{-4\nu}(1-\nu)+\cdots\ ,
\label{rsuno}
\\
R_{\mu\nu}R^{\mu\nu} &\approx & 
8G^2\, r^{-8\nu}\left[ (1-\nu)^2+\nu^2\right]+\cdots\ .
\label{rsdos}
\eea
The Ricci-squared scalar is nonvanishing for any real $\nu$.
Thus $\nu=0$ is a necessary condition for regularity but this is excluded by the (strong field) causality condition $\nu\geq \tfrac12$.
We thus see in a very direct way that strong-field causality alone  rules out the existence of  regular spherically-symmetric black-hole solutions of the Einstein-NLED equations.

In the literature the strong-field causality condition was overlooked.
If this condition is ignored, then one can remove the singularities in
$R$ and $R_{\mu\nu}R^{\mu\nu}$  by considering
theories with $\nu=0$ black holes.
 Even in these cases,
regularity also requires  finiteness of the Weyl-tensor squared at the origin.
This requires fine tuning of the parameters so that $M=\EE_{\rm em}$, in which case the $1/r$ term in $g_{tt}$ is absent.


\section{Exact Reissner-Nordstr\"om dyons}\label{sec:RNDyons}

Although the focus of this work is on implications of causality for black-hole solutions of 
the Einstein-NLED equations, there are many NLED theories that fail to be causal only for
sufficiently strong fields; i.e. they satisfy the weak-field causality/convexity conditions but fail to satisfy the strong-field causality condition \cite{Schellstede:2016zue,Russo:2024kto}. Such a theory may have utility as a weak-field effective field theory (EFT) because strong-field causality violations may then be indicative of a breakdown of the EFT approximation; the Euler-Heisenberg EFT for QED is a likely example. 

It is therefore of some interest to ask what can happen if we insist only on weak-field causality. 
For self-dual NLED theories (with a weak-field expansion) strong-field causality is implied by weak-field causality, so we consider theories that are not self-dual. The original Born theory is an example; its Lagrangian function 
is\footnote{In contrast to the Born-Infeld Lagrangian $L_{\rm BI} = T- \sqrt{T^2-2TS-P^2}$, this depends only on $S$.}
\be
L_{\rm Born} = T- \sqrt{T^2-2TS}\, . 
\label{bornL}
\ee
The Hamiltonian function is
\be\label{HBorn}
H_{\rm Born}= \sqrt{(T+ 2{\rm x})(T+2{\rm y})} -T \, , 
\ee
and this yields the following charged black-hole function:
\be
{\rm H}_{\rm Born}(r) = \sqrt{\left(T+ \frac{q_e^2}{r^4}\right) \left(T+ \frac{q_m^2}{r^4}\right)} -T\, . 
\ee
This function depends separately on $q_e$ and $q_m$ (rather than $Q$) but there is a $Z_2$-duality
invariance that exchanges them; this symmetry implies that the purely electric and purely magnetic black holes are isometric, with 
\be
{\rm H}_{\rm Born}(r)=   T\left\{ \sqrt{1+ \frac{Q^2}{Tr^4}} -1\right\}\, \qquad (q_eq_m=0),  
\ee
This is identical to the corresponding function for Born-Infeld, as expected because the Born and BI Hamiltonian functions coincide for purely electric or purely magnetic fields.

This coincidence of BI and Born black hole solutions for $q_eq_m=0$ does not apply to dyonic ($q_e q_m \ne0$) black holes. Consider, for example, the following special case: 
\be
q_e=q_m= Q/\sqrt{2}\,  \quad \rightarrow \quad {\rm H}_{\rm Born}(r)= \frac {Q^2}{2r^4} \, . 
\ee
This yields the RN metric! Thus, {\sl the RN dyonic black hole with $q_e=q_m$ is an exact solution of the Einstein-Born field equations, but {\sl not} of the Einstein-BI field equations}. 

The Born theory is not the only NLED theory, causal for weak fields,  for which the Einstein-NLED field equations have exact dyonic RN black-hole solutions. As we show in the following section, there is a class of such NLED theories with Lagrangian functions $L(S)$ (i.e. dependent on $S$ but not on $P$). We called them ``Born-type'' theories in \cite{Russo:2024kto} because Born's original theory, with $L= T- \sqrt{T^2-2TS}$ is a simple example and all (except Maxwell) violate causality for sufficiently strong fields \cite{Schellstede:2016zue}. In the following section we explain how for every Born-type theory the Einstein-NLED equations admit an exact dyonic Reissner-Nordstr\"om black-hole solution with $q_e=q_m$. 

\subsection{Born-type NLED} 

Consider the class of NLED theories defined by the Lagrangian function
\be
\hat L(S,\xi) = \xi S - \Omega(\xi)\, , 
\ee
where $\xi$ is an auxiliary scalar field that we restrict to be positive. We also restrict the function $\Omega(\xi)$ to be strictly convex ($\Omega''>0$);  this ensures uniqueness of any solution $\xi(S)$ to the auxiliary-field equation 
\be
\Omega' (\xi)= S\, . 
\ee
Thus, elimination of $\xi$ by its field equation yields a unique Lagrangian function
\be
L(S) := \hat L\left(S;\xi(S) \right)\, ,  
\ee
which has the following properties:
\be
L_S = \xi(S) >0\, , \qquad L_{SS} = \left[\Omega''\left(\xi(S)\right)\right]^{-1} >0 \, . 
\ee
Although no NLED theory defined by a Lagrangian function $L(S)$ is causal for sufficiently strong fields (with $P\ne0$) the above two properties ensure that it is causal for sufficiently weak fields \cite{Schellstede:2016zue,Russo:2024kto}. 

It is convenient to introduce a new unconstrained auxiliary field $\varphi$ by setting $\xi= e^\varphi$. We then have
\be
\hat L(S,e^\varphi) = e^\varphi S - V(\varphi) \, , \qquad V(\varphi) := \Omega( e^\varphi)\, . 
\ee

Since $\Omega'(e^\varphi) = e^{-\varphi}V'(\varphi)$, the existence of a stationary point of 
$\Omega(e^\varphi)$ implies a corresponding stationary point of $V(\varphi)$, which is unique when 
$\Omega$ is a strictly convex function; this condition translates to the following inequality for $V$:
\be
V''(\varphi) > V'(\varphi) \, . 
\ee 
This implies that $V''(\varphi)>0$ when $V'(\varphi) =0$; i.e. at the (unique, if it exists) stationary point of $V$, which is therefore a minimum of $V$. 

As an example, consider the potential function\footnote{The same choice yields Born-Infeld in the very different context of causal self-dual NLED \cite{Russo:2025fuc}.} 
\be
V(\varphi) = T\left\{\cosh\varphi -1\right\}\, . 
\ee
Elimination of the auxiliary field $\varphi$ yields
\be
L(S) = T- \sqrt{T^2-2TS}\, , 
\ee
which is the Lagrangian function of Born's original NLED theory. More generally, we shall say that any NLED theory with Lagrangian function $L(S)$ is of ``Born-type'' when it is causal for weak fields (i.e. $L(S)$ satisfies $L_S>0$ and $L_{SS}\geq 0$). We should caution here against confusion of this term with the ``Born-Infeld-type'' NLED theory that we will introduce later. 

Because the auxiliary-field form of the Born-type Lagrangian function is linear in $S$, the corresponding Hamiltonian function is easily found to be
\be\label{hatH}
\hat H({\rm x},{\rm y},\varphi)= e^{-\varphi} {\rm x} + e^{\varphi} {\rm y} + V(\varphi)\, , 
\ee
The auxiliary field can now be eliminated by means of the equation 
\be\label{auxH}
0= \frac{\partial \hat H}{\partial \varphi}  = - e^{-\varphi}{\rm x} + e^\varphi {\rm y} + 
V'(\varphi) \, . 
\ee
Evidently, this yields a Hamiltonian function $H({\rm x},{\rm y})$ that does not depend on the variable ${\rm z}$. The Lorentz-invariance condition of \eqref{LIH} reduces, for such functions, to
\be\label{LIz0}
H_{\rm x}H_{\rm y} =1\, . 
\ee
It is a simple matter to see that this is true; this Lorentz-invariance equation is formally the same as the one found in \cite{Russo:2024ptw} for Lorentz invariance of the Hamiltonian function of a self-dual NLED, and the auxiliary-field solution to it presented in \cite{Russo:2025fuc} yields \eqref{hatH} in the current context. Thus, \eqref{hatH} provides a Hamiltonian formulation for Born-type NLED 
theories. 

Notice that $H({\rm x},{\rm y})$ is symmetric under the interchange ${\rm x}\leftrightarrow {\rm y}$ when $V(\varphi)$ is an even function of $\varphi$ because this interchange can then be undone in $\hat H$ by a change of sign of $\varphi$ without this changing $V(\varphi)$. Within this `symmetric' subclass of Born-type theories (which includes the Born theory itself) the function $V'(\varphi)$ is odd in $\varphi$ and hence $V'(0)=0$. This shows that $\varphi=0$ solves \eqref{auxH} when ${\rm x}={\rm y}$, and for 
such solutions
\be
H = 2{\rm x} \qquad \rightarrow \quad {\rm H}(r) = \frac{q^2}{r^4} \qquad (q= q_e=q_m)\, .  
\ee
As for Born's theory, this leads to the RN metric. 

All that was necessary for this result was that $V'(0)=0$, and this can be true even when $V(\varphi)$ 
is not a symmetric function of $\varphi$, but then the stationary point of $V(\varphi)$ will typically be at $\varphi=\ \varphi_0$ for $\varphi_0\ne0$, and this leads to $L= e^{\varphi_0} S$ in the (Maxwell) weak-field limit. However,  our results for Einstein-NLED black holes use conventions for which $L=S$ for the Maxwell weak-field limit, so we should use for this purpose the rescaled Lagrangian function
\be
\tilde L= e^{-\varphi_0}L\, . 
\ee
In such cases, however, we can also define
\be
\tilde\varphi = \varphi - \varphi_0, \qquad \tilde V(\tilde\varphi) = e^{-\varphi_0}V(\varphi)\, , 
\ee 
to get 
\be 
\tilde L = e^{\tilde\varphi} S   - \tilde V(\tilde\varphi)\, , 
\ee
where, now, $\tilde V'(0)=0$. Thus, for fixed conventions for the Maxwell weak-field limit in the 
context of the full Einstein-NLED equations, we may assume without loss of generality that $V'(0)=0$.

In conclusion, the  familiar dyonic Reissner-Nordstr\"om solution of the Einstein-Maxwell equations with $q_e=q_m$ is {\sl also} a solution of the Einstein-NLED equations for {\it any} Born-type theory, i.e. any theory with Lagrangian $L=L(S)$ that satisfies $L_S>0$ and $L_{SS}\geq 0$. However, in the latter context we can expect that the failure of the Born-type theories to satisfy the strong-field causality condition will lead to instabilities of the RN dyon. In the following subsection we shall verify this for Born's original theory. 

\subsection{Instabilities}\label{subsec:stability}

We have seen that dyonic Reissner-Nordstr\" om black hole solutions of the Einstein-Maxwell equations
may also solve the Einstein-NLED equations when the NLED theory is of Born-type. However, the stability 
properties of any specific black-hole solution will generally depend on the particular field equations that it solves. We should restrict attention to NLED theories satisfying all weak-field causality 
conditions (i.e. for fields with energy densities much less than the Born tension $T$) because otherwise we can expect instabilities in the asymptotic region, arbitrarily far from the black hole horizon. This weak-field causality constraint is satisfied for Born-type theories (by our definition of them) but 
all NLED theories in this class are acausal for sufficiently strong fields \cite{Schellstede:2016zue}. 
In particular, the original Born theory is acausal in dyonic backgrounds with $|D||B|>T$ 
\cite{Russo:2024kto}. We shall show here that the $q_e=q_m$ RN black-hole is unstable as a solution of the Einstein-Born equations because of this strong-field acausality of Born electrodynamics. 

Results on (a)causality properties of NLED theories in a Minkowski spacetime background cannot be 
applied without modification when the spacetime metric is that of an RN black hole. However, let us consider the wave-propagation of some perturbation of the radial electric and magnetic fields of a  spherically-symmetric, and hence static,  charged black hole within a spherical shell bounded by the spheres $r=r_0\pm\epsilon$ for $r_0\gg r_H$ and $\epsilon \ll r_0-r_H$.  A small-amplitude but high frequency wave-packet centered 
at $r=r_0$ and propagating in a direction tangent to the radial unit vector will remain within the shell for long enough for it to be approximated as a small-amplitude plane wave moving in the constant static electromagnetic background defined by the fields at $r=r_0$.  In this approximation the spacetime metric within some small volume of the 
shell centered around a choice of radial vector of length $r_0$ is, approximately, 
\be
ds^2\approx -\NN_0^2 dt^2 + \NN_0^{-2} dr^2 + r_0^2 d\ell^2(\bb{E}^2)\,  \qquad \left[\NN_0 = \NN(r_0)\right]. 
\ee 
This is a Minkowski metric but with rescalings of the coordinates. 

Let us consider the RN solution of the Born theory for $q_e=q_m=q$. A small volume 
of the shell at radius $r=r_0$ in a fixed direction with unit vector ${\bf n}$ will 
have approximately constant and uniform vector densities $(D,B)$ in  the direction of 
${\bf n}$ with scalar magnitudes $q/r_0^2$. We may now use the Minkowski spacetime results of \cite{Mezincescu:2023zny} to find the dispersion relations of small amplitude plane waves in this background for wave vectors ${\bf k}$ orthogonal to ${\bf n}$. Taking into account the rescaling of the Minkowski coordinates one finds the following two dispersion relations for the two polarisations ($\pm$) of waves with angular frequencies $\omega_\pm$ and wave-vector magnitude $k$:
\be
\omega_+^2 = \NN_0^2 k^2\, , \qquad 
\omega_-^2 = \NN_0^2 \left[\frac{Tr_0^4-\NN_0^{-2} q^2}{Tr_0^4 + \NN_0^{-2} q^2}\right]
k^2\, . 
\label{frew}
\ee
These expressions may be compared with the Minkowski spacetime result obtained in \cite{Russo:2024kto}. For Born's theory, a wave propagating in a uniform, static electromagnetic background, has the dispersion relation:
\be
\omega_+^2 \bigg|_{\rm Mink} = k^2\ ,\qquad \omega_-^2\bigg|_{\rm Mink} = \left[\frac{T^2-B^2D^2}{(T+B^2)(T+D^2)}\right] k^2\, . 
\ee
where $|B|$ and $|D|$ are here the Minkowski spacetime scalar magnitudes of the constant 
uniform electromagnetic background.  We see that these formulas agree with \eqref{frew} in the limit  $\NN_0^2\to 1$, upon setting $B^2=D^2=q^2/r_0^4$.

Returning to \eqref{frew}, we see that $\omega_-^2\ge0$ requires 
\be
r_0^2\NN_0 \ge \frac{q}{\sqrt{T}}\, . 
\ee 
For simplicity, let us focus on the extreme RN metric, for which  $\NN = 1- GM/r$. 
Since $q= Q/\sqrt{2}$ in the current context, and $Q=\sqrt{\frac{G}{4\pi}}M$ for the extremal RN black hole, 
the above inequality simplifies to
\be
r_0^2- GMr_0 - \frac{GM}{\sqrt{(8\pi G) T}} \ge 0 \, ,  
\ee
with equality for the critical value $r_c$ of $r_0$ at which $\omega_-=0$.
This critical value is 
\be\label{rc}
r_c= \frac12 GM\left[ 1+ \sqrt{1+ \frac{4}{GM\sqrt{(8\pi G )T}}}\ \right]\, . 
\ee
For $r_0<r_c$ we have $\omega_-^2<0$, which indicates an instability of the static electric/magnetic fields against non-spherical perturbations. 

As $r_c\to GM$ in the $T\to\infty$ limit there is no instability outside the horizon, at $r=GM$, as expected because the $T\to\infty$ limit of Born is Maxwell. However, for finite $T$ there {\sl is} an instability with $r_c>GM$, i.e. not ``hidden'' behind the horizon as one might have thought possible. It is an instability against certain perturbations of the radial fields away from spherical symmetry. This conclusion assumes the validity of the approximations made in arriving at the formula \eqref{rc}. We expect these approximations to be valid when $r_c\gg GM$, which occurs when the ``gravitational Born length'' $1/\sqrt{GT}$ is much greater than the horizon radius; equivalently
\be
\ell_{gB} \gg \sqrt{G} Q\, . 
\ee
Comparison with \eqref{strong} shows that the RN black hole is one with very strong fields, but we think it likely that a less severe limitation on the 
result would emerge from a better calculation using the spherical symmetry 
of the background.

\section{Causality and Global structure}\label{sec:global}

We now turn to an analysis  of how causality restricts the possible global structures of static Einstein-NLED black-hole spacetimes. 
One important issue is the number of Killing horizons of the timelike vector field ($\partial_t$). These are hypersurfaces of constant $r$, with the constants given by the positive zeros of the function $g_{tt}(r)$. The Killing-horizon equation is therefore
\be\label{primera}
0=1-\frac{2G[M- \EE(r)]}{r} \, . 
\ee
For $r\ne0$ we  may rewrite this as 
\be\label{KHE}
2GM -r = 2G\EE(r)\, ,  
\ee
but we should remember that only $r>0$ solutions correspond to Killing horizons. 

\begin{figure}[h!]
 \centering
 \begin{tabular}{ccc}
\includegraphics[width=0.3\textwidth]{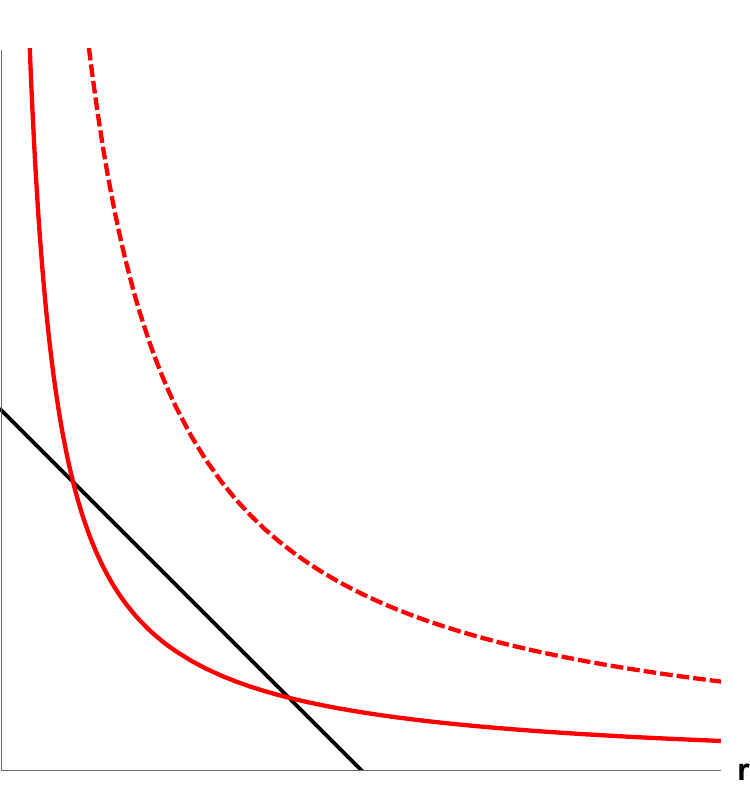}
 &
 \quad \includegraphics[width=0.3\textwidth]{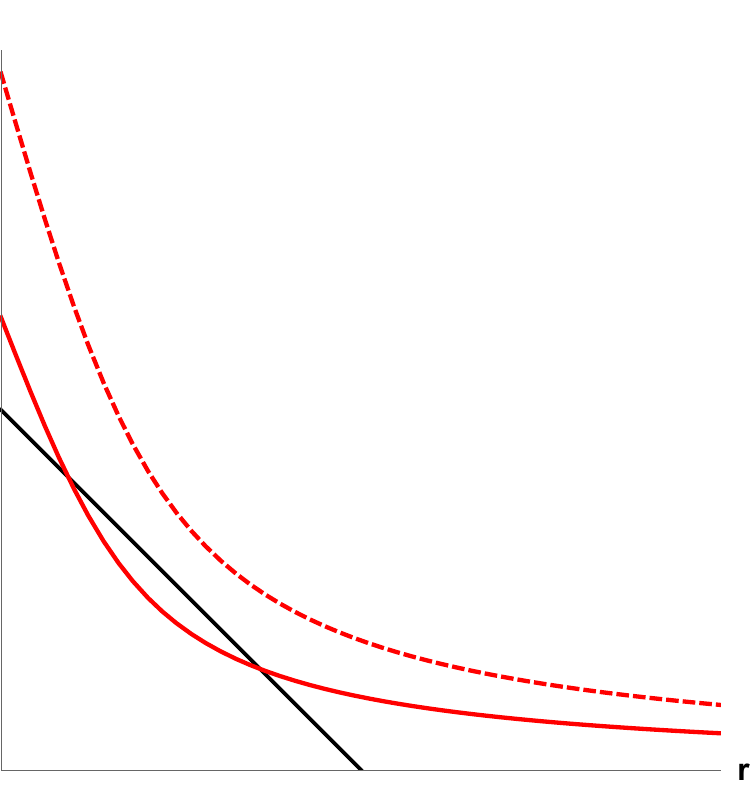}
  &
 \quad \includegraphics[width=0.3\textwidth]{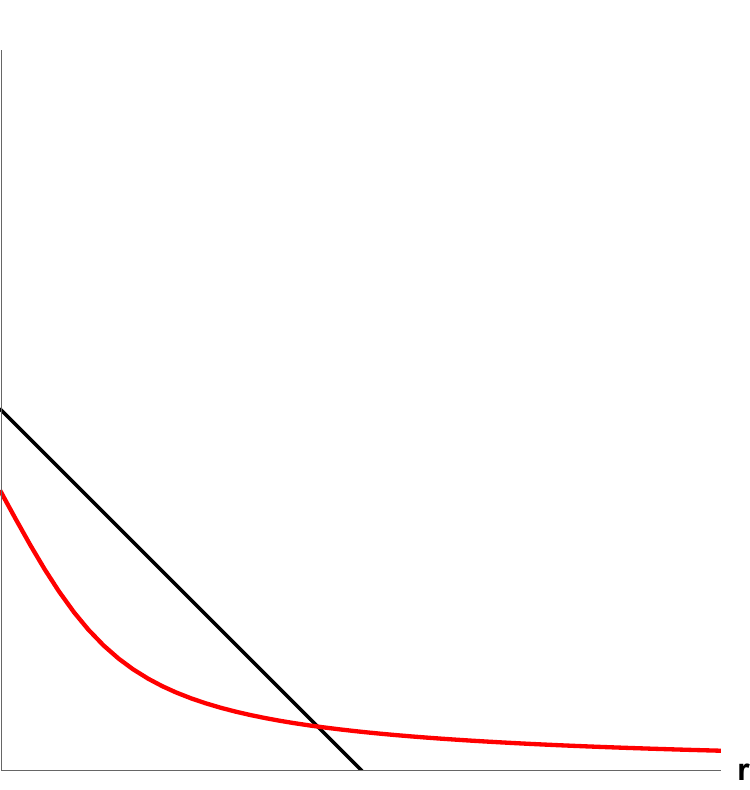}
 \\ (a)&(b)&(c)
 \end{tabular}
 \caption
 {Plot of the graphs of $2MG-r$ (black) and $2G\EE(r)$ (red). a) The electromagnetic energy 
 is infinite (e.g. Maxwell) and b) it is finite but $M< \EE_{\rm em}$. In these cases, charged black holes have Reissner-Nordstr\"om-like geometry: above extremality (solid red) there are two Killing horizons (event horizon and Cauchy horizon) and below extremality (dashed) there is no horizon (timelike naked singularity). c) $M>\EE_{\rm em}$: one Killing horizon (the event horizon of a charged black hole with Schwarzschild-like geometry).  
 }
 \label{eeab}
 \end{figure}

In those cases, such as Maxwell, for which $\EE(0)$ is infinite (i.e. infinite electromagnetic energy) the graph of the right-hand side of  \eqref{KHE} will intersect the straight line graph of the left-hand side an even number of times (counting osculation points as double intersections), but the strict convexity\footnote{Convexity alone does not suffice since this would allow red line in fig. \ref{eeab}a to coincide with some portion of the black line. 
} of $\EE(r)$ restricts the number of intersections to either $0$ (no horizons, and hence a naked singularity) or $2$ (an event horizon and Cauchy horizon, as for RN black holes, which combine to 
form the degenerate horizon of an extreme black hole when they coincide); this is illustrated in fig. \ref{eeab}a. For any causal NLED theory for which charged black holes have infinite electromagnetic energy, the possible global structures are those of the RN family of charged black holes of the Einstein-Maxwell equations. These are ``RN-like'' black holes.

The above argument is similar to one given in \cite{Hale:2025ezt} on the assumption of finite electromagnetic energy: $\EE(0) = \EE_{\rm em}$. For $M< \EE_{\rm em}$ the analysis is essentially the same as for infinite electromagnetic energy (as shown in fig. \ref{eeab}b) but for $M\ge \EE_{\rm em}$ the graph of $2G\EE(r)$ lies below the straight-line graph of $(2GM-r)$ for small $r$ and will therefore intersect this line precisely once (as shown in fig. \ref{eeab}c). The value of $r$ at this intersection is the radius of the event horizon of a charged black hole without a Cauchy horizon. 
Therefore, {\sl finite electromagnetic energy implies that all charged black holes with 
$M>\EE_{\rm em}$ are Schwarzschild-like (S-like) with an interior spacelike singularity 
at $r=0$}, and all RN-like black holes have mass $M<\EE_{\rm em}$. This statement was proved by Hale et al. \cite{Hale:2025ezt} but in the context of a subset of charged black holes with finite electromagnetic energy, those of ``BI-type'' ($\nu=\frac12$) for which the small-$r$ expansion of $\EE(r)$ is as they assumed it to be. We now see that it is a consequence of causality for {\sl all} charged black holes with finite electromagnetic energy ($\frac12\le \nu<\frac34$), although the BI-type subset is special in other respects. 

We have still to consider $M= \EE_{\rm em}$, which we have shown above to be the boundary of the region in the $(M,Q)$ parameter space corresponding to S-like black holes. On dimensional grounds, the value of $\EE_{\rm em}$ must take the same form as the BI result of \eqref{BIem}, i.e. 
\be\label{S-boundary}
\EE_{\rm em}  = C T^\frac14 Q^{\frac32}\, , 
\ee 
for positive numerical coefficient $C$, which may also be a function of dimensionless parameters such as the $q_e/q_m$ ratio. The (non-negative) values of $(M,Q)$ for which $M= \EE_{\rm em}$ are points on a curve in this parameter space. Here we shall use coordinates $(m,Q)$ for this space, where $m$ is the following dimensionless version of the mass-to-charge ratio:
\be\label{M/Q}
m :=  \sqrt{\frac{G}{4\pi}} \left(\frac{M}{Q}\right)\, . 
\ee 
In these coordinates, the curve $M= \EE_{\rm em}$ is the following half-parabola:
\be 
m^2 = \left(\frac{C^2 G\sqrt{T}}{4\pi}\right)\, Q \qquad \qquad (M= \EE_{\rm em})\, .  
\ee 
This divides the (physically relevant) positive quadrant of the $(m,Q)$ plane into an upper part corresponding to S-like black holes and a lower part corresponding to RN-like black holes and naked timelike singularities (with their distribution yet to be determined). 

Whenever the electromagnetic energy is finite we may use 
\eqref{EE-finite} to put the Killing-horizon equation of \eqref{KHE} into the form 
\be\label{key}
2G(M-\EE_{\rm em}) = r\, -\, 8\pi G\int_0^r d\rho \left[\rho^2 {\rm H}(\rho)\right]\, . 
\ee
Notice that the right-hand side  goes to zero as $r\to 0$. This tells us that any limit in which a Killing horizon radius 
goes to zero is a limit for which $M\to \EE_{\rm em}$. This could be the radius $r_-$ of a Cauchy horizon that moves to 
$r=0$ as the mass of an RN-like black hole is increased at fixed $Q$ until $M= \EE_{\rm em}$; in this case the event horizon radius $r_+$ remains non-zero. It could also be the event horizon radius of an S-like black hole as its mass is decreased until $M=\EE_{\rm em}$, but this can only happen in a direct transition from S-like black hole to naked timelike singularity. As we shall see in the following subsection, the first of these possibilities is generic whereas the second possibility is realised only in certain (BI-type) causal NLED theories for certain (small-charge) black holes. 

\subsection{The S-like boundary}\label{subsec:SBoundary}

On the S-like boundary the Killing-horizon equation of \eqref{key} simplifies to 
\be\label{preF}
r= 8\pi G \int_0^r  d\rho \left[\rho^2 {\rm H}(\rho)\right]\, , \qquad \qquad (M= \EE_{\rm em}).
\ee 
We now investigate the implications of this equation near $r=0$ for the `minimal' causal  family of self-dual NLED theories (with parameter $\nu$) defined by \eqref{frakh1}. The small-$r$ expansion of 
$r^2{\rm H}(r)$ for this family is given in \eqref{frakh3}, and using it we find the following equation 
\be\label{gen-nu}
r= 8\pi G T \left(\frac{Q^2}{2\nu T}\right)^\nu r^{3-4\nu} \left\{ \frac{1}{3-4\nu} - \frac13\left(\frac{Q^2}{2\nu Tr^4}\right)^{-\nu} + \ \mathcal{O}\left(\frac{Tr^4}{Q^2}\right) \right\}\, . 
\ee 
The $\nu=\frac12$ case (Born-Infeld) is special because the leading term on the right-hand side (as $r\to 0$) is (like the left-hand side) linear in $r$. For all $\nu>\frac12$ (but $\nu<\frac34$) it tends to zero more slowly than $r$ (whereas all other terms tend to zero faster than $r$, and can be neglected for sufficiently small $r$). We therefore need a separate analysis for $\nu=\frac12$ and $\frac34> \nu >\frac12$:
\begin{itemize}

    \item $\nu= \frac12$. The terms linear in $r$ can be combined. We then find (for $r\ne0$ and dividing by $r$) that 
\be
    \left(Q-Q^{\rm BI}_{\rm cr}\right) = 
    \frac13 \sqrt{T} r^2 \left\{ 1 + \mathcal{O}\left(\sqrt{T}r^2/Q\right)\right\}\, , 
\ee
where $Q_{\rm cr}^{\rm BI}$ is a ``critical charge'', in this Born-Infeld case, 
\be\label{Qcrit}
Q^{\rm BI}_{\rm cr} = \frac{1}{8\pi G\sqrt{T}}\, . 
\ee 
For small positive $(Q-Q^{\rm BI}_{\rm cr})$ there is a real positive solution for $r$. We shall see later that this is the event horizon of a charged black hole that lies on a transition point between S-like ($M>\EE_{\rm em}$) and RN-like ($M<\EE_{\rm em}$) black holes. When 
$Q<Q^{\rm BI}_{\rm cr}$ there is no real positive solution for $r$ near $r=0$
and the central singularity is naked. 

\item $\frac34> \nu>\frac12$. In this case, we can proceed by assuming that $r^4 \ll Q^2/T$; as pointed out in subsection \ref{subsec:NSE} this restricts $r$ to be much less than the geometric mean of $\ell_{gB}$ and 
$\sqrt{G}Q$. In this case \eqref{gen-nu} reduces to 
\be
r^{2\nu-1} \approx \sqrt{\frac{8\pi GT}{3-4\nu}} \left(\frac{Q^2}{2\nu T}\right)^{\nu/2}\, . 
\ee
For sufficiently small $r$ (which it is when $\ell_{gB} \gg \sqrt{G} Q$) we again have one real positive solution,  but now for all $Q$ because there is no critical charge. 

\end{itemize}
Numerical factors (depending on additional dimensionless parameters) will differ for other models, but these will not change the crucial difference between the leading terms of \eqref{gen-nu} for $r\to 0$ for the $\nu=\frac12$ and $\frac34> \nu>\frac12$ cases. In particular, 
there is a critical charge iff $\nu=\frac12$, as we confirm in the following subsection. 

To summarise; for $\frac34> \nu>\frac12$, every point on the $M= \EE_{\rm em}$ half-parabola in the $(m,Q)$-plane corresponds to an S-like black holes that can also be viewed as a limit of an RN-like black hole in which the Cauchy horizon 
contracts to $r=0$. For $\nu=\frac12$, points on the half-parabola are S-like black holes with the same interpretation provided $Q$ exceeds some finite critical value $Q_{\rm cr}$, but for $Q< Q_{\rm cr}$ they are naked singularities arising from the $r_h\to0$ limit (zero-entropy limit in the quantum theory) of an S-like black hole. 

In both cases RN-like black holes are possible, but there is a minimal mass $M_{\rm ext}$ for fixed charge, which is reached when the Cauchy horizon expands to coincide with the event horizon to form a degenerate (zero surface gravity) 
event horizon. For any charge $Q$ that allows RN-like black holes ($Q>0$ generically but $Q>Q_{\rm cr}$ for BI-type) any spacetime with $M<M_{\rm ext}$ has a naked timetime singularity at $r=0$. Therefore, the extreme  black holes form an additional boundary to the region of the $(m,Q)$-plane corresponding to RN-like black holes. 

The type of phase diagram (delineating regions in parameter space of S-like and RN-like black holes from each other and from spacetimes with naked timelike singularities) depends on the locus of the extreme-black-hole boundary curve in relation to the S-like boundary curve. We begin an investigation of this relation in the following subsection.

\subsection{Extreme black holes}\label{subsec:ECBH}

For an RN-like black hole there are two Killing horizons ($r= r_\pm$): an event horizon at $r=r_+\ge r_-$
and a Cauchy horizon at $r_->0$. In the case that $r_+=r_- =r_h$ we have a single degenerate (zero surface gravity) Killing horizon, which is now the event horizon of an extreme black hole with $M= M_{\rm ext}$, the extremal mass. Extremality ($r_\pm = r_h$) implies that $g_{tt}(r)$ has a  double zero at $r=r_h$, so its derivative with respect to $r$ must also be zero at $r=r_h$. This gives us the following two equations for extremal black holes:
\be
\label{horieq}
2GM = r + 2G\EE(r) \, , \qquad 0 = 1 + 2G\EE'(r)\, . 
\ee
The first of these equations, which is equivalent to \eqref{KHE},  determines $M$ as a strictly-convex function of $r$ (assuming causality) which therefore has either no stationary points or one stationary point, an absolute minimum. The second, ``extremality'', equation tells us that $r$ at this stationary point (if it exists) is the event horizon radius $r_h$ of an extremal black hole, and $M$ is then the extremal mass $M_{\rm ext}$. Therefore 
\be\label{Mext}
M_{\rm ext} = \frac{r_h}{2G} + \EE(r_h) \, , \qquad F(r_h) =1\, , 
\ee
where 
\be\label{Fdef}
F(r) := -2G\EE'(r) \equiv 8\pi G \, r^2 {\rm H}(r) \, . 
\ee 
Since the negative function $-2G\EE(r)$ is strictly concave and approaches zero as $r\to\infty$, the positive function $F(r)$ is a monotonically decreasing function of $r$, which also approaches zero as $r\to\infty$. It follows that 
\be\label{Fineq}
\int_0^r F(\rho) d\rho\   > \ rF(r)\, .  
\ee
Using \eqref{Fdef}, we may now rewrite \eqref{key} as
\be\label{key2}
2G(M-\EE_{\rm em}) = r- \int_0^r \! F(\rho) d\rho \, < \, \left[1-F(r)\right]r\, , 
\ee
where the inequality follows from \eqref{Fineq}. This is true for any Killing horizon radius $r$;
if we now choose it to be the event-horizon radius of an extremal black hole, with $M= M_{\rm ext}$ and $F(r_h)=1$, 
we deduce that 
\be
M_{\rm ext} < \EE_{\rm em}\, . 
\ee
This should not be a surprise because $M\ge M_{\rm ext}$ for RN-like black holes at fixed $Q$, and we have already seen that all RN-like black holes with finite electromagnetic energy have $M<\EE_{\rm em}$. The RN-like black holes correspond to a region in parameter space that is below the parabolic S-like boundary curve in the $(m,Q)$-plane but above the curve on which $M=M_{\rm ext}$. These two curves divide the parameter space into three regions, one for each of the three phases, but all three phases will be possible for all $Q>0$ only if the two boundary curves never meet (i.e. have no point in common) except at $Q=0$. However, it may happen they meet for $Q>0$, and in this case there will be a restriction on the possible values of $Q$ for RN-like black holes, in particular extreme black holes.

To investigate this point we return to the extremality equation $F(r_h)=1$, which has at most one solution; this is because there cannot be more than two Killing horizons, but it is also a direct consequence of the fact that $F(r)$ decreases monotonically to zero as $r\to\infty$. So, either there is an extreme black hole or there is not, and which of these two possibilities is realised depends on the behaviour of $F(r)$ as $r\to 0$. We must now distinguish between the following cases:

\begin{enumerate}
    
\item $F(r) \to\infty$ as $r\to 0$. In this case the equation $F(r)=1$ always has a (unique) solution $r=r_H>0$; i.e. at the (degenerate) event horizon of an extremal charged black hole. 

\item $F(0)$ is finite. There are now two subcases
\begin{itemize}
\item $F(0)>1$. The equation $F(r)=1$ has a unique solution at the event-horizon radius of an extreme black hole.
\item $F(0)\le1$. The equation $F(r)=1$ has no solution with $r>0$; there is no extremal black hole.
\end{itemize}
\end{enumerate}
We can explore these possibilities further by assuming the power-law behaviour of \eqref{Hrpower} as $r\to 0$. We may then use \eqref{Esim} to conclude that 
\be
F(r) \sim r^{2(1-2\nu)} \qquad (r\to 0)\, . 
\ee
This shows that $F(r)$ diverges as $r\to0$ for all $\nu>\frac12$. Item 1 above therefore applies, and a unique extreme charged black hole  exists for any value of the charge $Q$. 

In contrast, $F(0)$ is finite for $\nu=\frac12$ as $r\to0$, and item 2 applies.
In this BI-type case $H\sim r^{-2}$. On dimensional grounds
$$
H\, \sim\,  T \left(\frac{Q^2}{T}\right)^{\frac12}\ r^{-2}\ .
$$
Therefore
$$
F(0)=\lim_{r\to 0} 8\pi G r^2 H(r) =2G b_1 \sqrt{T}Q\ ,
$$
where $b_1$ is a positive numerical coefficient.
The existence of an extreme black hole therefore depends on which of the item 2 subcases applies.
The condition $F(0) >1$ for the existence of an extreme black hole is now equivalent to the condition $Q > Q_{\rm cr}$, where
\be
 Q_{\rm cr}= \frac{1}{ 2 G b_1 \sqrt{T}}\ .
\ee
 For Born-Infeld (for example) 
\be
F(r_h) = -8\pi G r_h^2 T\left(1-\sqrt{1+\frac{Q^2}{T r_h^4}}\right) \quad  \Rightarrow \quad F(0) = 8\pi G \sqrt{T} Q\, ,  
\ee
and therefore $Q_{\rm cr}=1/ (8\pi  G  \sqrt{T})$ (i.e. $b_1=4\pi$), in 
agreement with \eqref{Qcrit}.

What both the generic and BI-type cases have in common is that the lower RN-like boundary curve ($M=\EE_{\rm em}$) will meet the S-type boundary curve ($M=M_{\rm ext}$) whenever $r_h\to0$. The difference is that this occurs for zero $Q$ in the generic case and for positive $Q$ in the BI-type case. In both cases this meeting point is a ``triple-point'' for the three distinct `phases' of spherically symmetric spacetimes (RN-like and S-like black holes and naked timelike singularities): any neighbourhood of this point contains a subregion of the region for each of the three `phases'. However, 
all three phases are possible for arbitrarily small $Q$ only if the triple point occurs at $Q=0$, as is the case generically. For the BI-type cases, there are no RN-like black holes for $Q<Q_{\rm cr}$. 

An obvious question now, for both generic and BI-type cases, is whether the extreme black hole curve (the lower boundary for RN-like black holes) can meet the S-like boundary curve again, i.e. for positive $r_h$ and some larger value of $Q$. If $r>0$ is limiting value of $r_h$ on the $M=M_{\rm ext}$ boundary curve at such a new meeting point then it must be a solution of the following two equations:
\be\label{meetings}
r = \int_0^r \! F(\rho)\, d\rho  \, , \qquad  F(r)=1\, . 
\ee
However, since $F(r)$ is a monotonic decreasing function, we have (for $r>0$)
\be
\int_0^r \! F(\rho)\, d\rho \ >\  r F(r) = r\, , 
\ee
where we have used the extremality condition $F(r)=1$ for the final equality. Using this result in the first of the equations of \eqref{meetings}, we arrive at the contradiction $r>r$. Therefore, in all cases, the only meeting point is the one 
for which $r_h\to0$ that we considered initially.  This is significant because it implies that {\sl the qualitative structure of the phase-diagram is determined by its structure in a neighbourhood of the one triple point}. We analyse this structure in more detail in subsection \ref{subsec:Cauchy}. 

All (spherically symmetric) black holes of the Einstein-BI equations are BI-type black holes, and in this case we can solve the extremality $F(r_h) =1$ explicitly for the horizon radius $r_h$:
\be\label{BIcase}
r_h^2 = 4\pi G \left(Q^2- Q^2_{\rm cr}\right) \, \qquad ({\rm BI}). 
\ee 
This illustrates the point that there is only one triple point, at $Q=Q_{\rm cr}$, because if there were another one at a meeting point with a larger value of $Q$ then there would be no extreme black hole for that value of $Q$, in contradiction to the formula \eqref{BIcase}. Inspection of this formula provides additional details about the properties of extreme black holes, at least for Born-Infeld. 

For example, for large $Q$ we have 
\be
\lim_{Q\to\infty} \left[r_h/Q\right] = \sqrt{4\pi G} \, .
\ee
Similarly, in the $r_h\to\infty$ limit the extremality equation $F(r_h)=1$ approaches the equation one would find from the weak-field limit theory, which yields $r_h/Q= \sqrt{4\pi G}$. 
Thus,  {\sl $Q\to\infty$ is a weak-field limit}. 

Another feature of the BI formula \eqref{BIcase} is that the ratio $r_h(Q)/Q$ is not only asymptotic to $\sqrt{4\pi G}$ but also increases monotonically to this limiting value. 
As we now show, this is true for any causal NLED theory.

\subsubsection{Monotonicity of $r_h(Q)/Q$}

For black holes with arbitrary charges $(q_e,q_m)$, the black-hole function ${\rm H}(r)$ is found from the Hamiltonian function $H({\rm x},{\rm y},{\rm z})$ by using the expressions of \eqref{xyz} for the scalar variables $({\rm x},{\rm y},{\rm z})$; i.e,
\be
{\rm H}(r) = H\left(\frac{q_e^2}{2r^4}, \frac{q_m^2}{2r^4}, \frac{q_e q_m}{r^4}\right)\, . 
\ee
Since $Q^2= q_e^2+q_m^2$, we have, for some angle $\varphi$,
\be
q_e = Q \sin\varphi\, , \qquad  q_m = Q\cos\varphi \, , 
\ee
and therefore 
\be\label{Qtor}
Q\partial_Q {\rm H}(r) = -\frac12 r{\rm H}'(r) \, . 
\ee
Using \eqref{Fdef} we can rewrite the extremality equation $F(r_h)=1$ as
\be
r^2_h {\rm H}(r_h)=\frac{1}{8\pi G}\, . 
\ee
This equation implicitly determines $r_h$ as a function of $Q$: by taking the derivative respect to $Q$ on both sides 
and using \eqref{Qtor} we find that 
\be\label{drdQ}
\left[r_h {\rm H}'(r_h) + 2{\rm H}(r_h)\right] \frac{d r_h}{dQ} =  r_h {\rm H}'(r_h) \left(\frac{r_h}{2Q}\right)\, . 
\ee
This is equivalent to the equation 
\be\label{QdQln}
Q\frac{d}{dQ} \left[\ln \left(\frac{r_h}{Q}\right)\right] = -
\frac{2\pi r_h\,  \Theta(r_h)}{\EE''(r_h)} \, ,  
\ee
where we have used the expression for $\Theta(r)$ of \eqref{defTheta} and the identity for $\EE''(r)$ of \eqref{convexE}. 

As explained in section \ref{sec:ADM}, weak-field causality requires $\Theta\le0$ (with equality for conformal NLED theories) and strong-field causality requires $\EE''(r) >0$ (i.e. strict convexity). It follows that for all causal NLED theories
\be
Q\frac{d}{dQ} \left[\ln \left(\frac{r_h}{Q}\right)\right] \ge 0\, ,  
\ee
where equality for non-conformal theories is achieved only asymptotically as $Q\to\infty$. 

This inequality is equivalent to the statement that $r_h(Q)/Q$ 
is a monotonic increasing function of $Q$ (which implies the same for $r_h(Q)$). Since $r_h(Q)/Q$ tends to its constant free-field\footnote{We assume a Maxwell weak-field limit here, for simplicity.} value 
as $Q\to\infty$, for any finite $Q$ we have
\be
r_h(Q) \le \sqrt{4\pi G}\, Q\ ,
\ee
with equality only in the $Q\to\infty$ limit. 
In other words, the event horizon radius (and therefore its area) is less, for any given charge $Q$, than it is for the RN 
extreme black hole. 

To put it another way, causal NLED interactions reduce the entropy of an extreme black hole for fixed charge. This reduction is small for large $Q$ (since $Q\to\infty$ is a weak-field limit) but becomes increasingly large for decreasing $Q$.

\subsection{Extremal mass-to-charge ratio}

We introduced in \eqref{M/Q} the dimensionless parameter $m$ proportional to the 
mass-to-charge ratio $M/Q$ for any spherically-symmetric and asymptotically-flat solution of the Einstein-NLED equations. Applied to RN-like spacetimes, there is a minimal value of $M$ at fixed $Q$, the mass $M_{\rm ext}$ of the extreme RN black hole of charge $Q$, and therefore a minimal value of $m$ (again at fixed $Q$):
\be\label{mudef}
m\ge \mu := \frac{GM_{\rm ext}}{\sqrt{4\pi G}\, Q}\, . 
\ee 
For generic non-self-dual NLED, $\mu$ is both a theory-dependent function of $Q$ {\sl and} a function of the $q_e/q_m$ ratio. The normalization is such that $\mu=1$ for the extreme RN black hole, and since $Q\to\infty$ is a weak-field limit
(because $r_h(Q)\to\infty)$, we have 
\be
\lim_{Q\to\infty} \mu(Q) = 1 \, . 
\ee

It was shown in  \cite{Abe:2025vdj}, initially for magnetic black holes, 
that $\mu(Q)$ is a monotonic increasing function of $Q$; an extension of this result to dyons by means of a duality transformation was subsequently proposed but we have not fully understood the argument\footnote{For example, the duality transformation used in \cite{Abe:2025vdj} is claimed to be an invariance of the stress-energy tensor, but the Hamiltonian is a component of this tensor and its duality invariance is a standard definition of what is meant by self-duality.}. Here we 
present a simple general proof using a Hamiltonian approach.

We begin with the observation that
\be
r_h = -8\pi G\int_{r_h}^\infty \! dr \left[r^3 {\rm H}(r)\right]'\, , 
\ee 
which follows from the extremality equation $F(r_h)=1$, and the fact that $r^3{\rm H}(r) \sim Q^2/r$ as $r\to\infty$. This equation for $r_h$ is trivially rewritten as 
\be\label{rh2}
\frac{r_h}{2G}  =  -4\pi \int_{r_h}^\infty \! dr \left[ r^3 {\rm H}'(r)\right] \ - \ 3\, \EE(r_h)\, . 
\ee 

Now, using \eqref{Qtor} in \eqref{intEE}, we learn that
\be\label{QdQE}
Q\partial_Q \EE(r) = 
-2\pi \int_{r}^\infty \! dr \left[ r^3 {\rm H}'(r)\right]\, ,  
\ee 
which allows us to rewrite \eqref{rh2} as 
\be\label{later}
\frac{r_h}{2G} = 2Q \partial_Q \EE(r_h) -3  \EE(r_h)\, .  
\ee 
We shall need this equation below. 

Turning now to the definition of $\mu$ in \eqref{mudef}, we have 
\be
\left(\sqrt{\frac{4\pi}{G}}\right) Q^2\frac{d\mu}{dQ} \ = \  Q(dM_{\rm ext}/dQ) - M_{\rm ext} \, . 
\ee
On the right-hand side we now use the expression for $M_{\rm ext}$ in \eqref{Mext}. Since 
$M_{\rm ext}$ is the value of the function $[r/(2G)+ \EE(r)]$ at its absolute minimum (this is the extremality equation) a variation of $r_h$ induced by a variation of $Q$ has no first-order effect on $M_{\rm ext}$; therefore
\be
Q(dM_{\rm ext}/dQ) =  Q\partial_Q \EE(r_h)\, . 
\ee
We thus find that 
\bea
\left(\sqrt{\frac{4\pi}{G}}\right) Q^2\frac{d\mu}{dQ} \ &=& \   
Q\partial_Q\EE(r_h) - \EE(r_h) - \frac{r_h}{2G} \, \nonumber \\
&=& \ - \left(Q\partial_Q -2\right) \EE(r_h) \, , 
\eea
where we have used \eqref{later} for the second equality. Using \eqref{QdQE} we have
\bea\label{dmudQ}
-\left(Q\partial_Q -2\right) \EE(r_h) &=& 2\pi \int_{r_h}^\infty \! dr \ r^2 
\left[r{\rm H}'(r) + 4{\rm H}(r) \right] \nonumber \\
&=&\  - 2\pi \int_{r_h}^\infty\! dr\ r^2\Theta(r) \, , 
\eea
where we have used \eqref{defTheta} for the second equality.
Thus, 
\be
\label{dmuq}
Q^2\frac{d\mu}{dQ} \ = -\sqrt{\pi G} \int_{r_h}^\infty \! dr\ r^2\Theta(r) \, , 
\ee
in agreement with \cite{Abe:2025vdj}. Since $\Theta\le0$ for causal NLED theories, we have
\begin{equation}
    \frac{d\mu}{dQ} \ge 0\ ,
\end{equation}
with equality only for conformal NLED. 

In contrast to the proof of monotonicity of $r_h(Q)/Q$, this proof of monotonicity of $\mu(Q)$ uses only the {\sl weak-field} causality condition $\Theta \le 0$. It therefore holds for many theories that are not causal, such as the Born-type theories discussed in section \ref{sec:RNDyons}.


\subsubsection{Causality and the Weak Gravity Conjecture}

In the context of  higher-derivative  corrections to the Einstein-Maxwell theory, which affect the mass-to-charge  ratio of extreme  charged black holes, it was observed by Cheung et al. that the weak-gravity conjecture (WGC) requires this ratio to be a monotonically increasing function of the charge, because then “large black holes are unstable to decay  to smaller ones” \cite{Cheung:2018cwt}.  This observation was used in \cite{Abe:2025vdj} to interpret the monotonicity property of $\mu(Q)$ as a relation between the WGC and causality. Here we comment briefly on this issue, as we intend to return to it in a larger context in  later work. 

The decay of any extreme black hole of charge $Q=Q_1+Q_2$ into two extreme black holes of (positive) charges $Q_1$ and $Q_2$ is energetically possible iff the extremal mass function $M_{\rm ext}(Q)$ satisfies the  following ``superadditivity’’ condition:
\begin{equation}
\delta : = M_{\rm ext}(Q_1+Q_2) - M_{\rm ext}(Q_1) -  
M_{\rm ext}(Q_2) \ge 0 \, . 
\end{equation}
The proof that this is implied by the monotonicity property of $\mu(Q)$, which is implied by (``weak-field'')  causality, is simple: 
\begin{equation}
\sqrt{\frac{G}{4\pi}}\ \delta \equiv Q_1\left[\mu(Q_1+Q_2)-\mu(Q_1)\right]+Q_2\left[\mu(Q_1+Q_2)-\mu(Q_2)\right]\geq 0\ . 
\end{equation}
The final inequality follows from the fact that monotonicity requires both terms to be non-negative;  it is saturated only for conformal NLED.

\subsection{Triple points}\label{subsec:Cauchy}

In this subsection we focus on black holes with finite electromagnetic energy. Only in these cases are three phases possible, and this opens up the possibility of `triple points' at which all three phases meet. 
The fact (established in subsection \ref{subsec:ECBH}) that a triple point is also a point at which $r_h\to0$ (where
$r_h$ is the event-horizon radius of an extreme RN-like black hole) allows us to confirm crucial aspects of our general analysis by a detailed study of phase diagrams in a neighbourhood of a triple point by means of small-$r$ expansions.  As discussed briefly in subsection \ref{subsec:ECBH}, there are two qualitatively different triple points to consider:
\begin{itemize}

    \item The $M=\EE_{\rm em}$ and $M= M_{\rm ext}$ curves meet only at $Q=0$, and all points between them correspond to non-extremal RN-like black holes. All S-like black holes on the S-like boundary have non-zero horizon radius, and are limits of an RN-like black hole in which the Cauchy horizon has shrunk to $r=0$. There are RN-like black holes for any $Q>0$ because 
    the triple point is at $Q=0$. This possibility is generic in the sense that it applies whenever $\frac34 > \nu>\frac12$ (and for $1 \ge \nu \ge \frac34$ but in these cases the electromagnetic energy is not finite). 
    
    \item The $M=\EE_{\rm em}$ and $M= M_{\rm ext}$ curves meet only at $Q=Q_{\rm cr}>0$, which implies that all RN-like black holes have $Q>Q_{\rm cr}$. The triple point is now at $Q=Q_{\rm cr}$. The segment of the S-like boundary with 
    $Q <Q_{\rm cr}$ is now also an upper boundary for naked timelike singularities; points on this portion of the $M= \EE_{\rm em}$ curve can therefore be found as limits of S-like black holes for which the horizon radius has shrunk to zero (without a previous transition to an RN-like black hole). This possibility is realised only for BI-type black holes:  $\nu=\frac12$ (in particular, all those of the Einstein-BI equations).

    \end{itemize}
    
    We now present a more detailed analysis of these triple points, starting with the  BI-type black holes. 
For simplicity 
we focus on the one-parameter family of self-dual BI-type NLED theories defined by the Lagrangian CH-function of \eqref{arem}. 
The small-$r$ expansion of the $\EE(r)$ function is given in \eqref{aremE}; we may rewrite it as  
\be\label{BI+}
\EE(r)= \EE_{\rm em}  - b_1\sqrt{T}\, Q\, r  + b_2 T r^3 +
\mathcal{O}(r^\lambda) \, ,
\ee
where\footnote{The addition of a cosmological constant changes the $b_2$ coefficient but this has no effect on the singular term in the spacetime 
metric.}
\be
b_1 = 4\pi \sqrt{2\alpha} \, , \quad b_2 = \frac{4\pi}{3} \, ; \qquad
\lambda = \frac{3-\alpha}{1-\alpha} > 3\, . 
\ee 
On dimensional grounds, we expect \eqref{BI+} to apply to any $\nu=\frac12$ charged black hole in the context of causal NLED theories, irrespective of the charges $(q_e,q_m)$, 
modulo factors of positive dimensionless functions of any additional dimensionless parameters multiplying each term in the expansion. In what follows we shall therefore assume only that $(b_1,b_2)$ are positive constants. 

Using \eqref{BI+} in \eqref{primera}, we find the following small-$r$ expansion of $g_{tt}$:
 \be\label{desarr}
 -g_{tt} = 1- \frac{2G\left[M -\EE(r)\right]}{r} =  \frac{\epsilon}{r}  
 - \zeta + \frac{r^2}{r_0^2} + \mathcal{O}(r^4)\, , 
 \ee
 where\footnote{Notice that $\zeta$ is dimensionless while $(\epsilon,r_0)$ have 
 dimensions of length.} 
 \be\label{ep}
 \epsilon : = 2G\left(\EE_{\rm em}-M\right)\, , \qquad 
 r_0 := \frac{1}{\sqrt{2b_2 GT}}\  \sim \ \ell_{gB}\, ,  
\ee
and 
\be\label{zet}
\zeta := \frac{Q-Q_{\rm cr}}{Q_{\rm cr}} \, , \qquad \sqrt{G}Q_{\rm cr} = \frac{1}{2b_1 \sqrt{GT}} \ \sim \ \ell_{gB}\, . 
\ee
The equation for the Killing-horizon radii is now 
\be\label{cubic}
r^3 - (\zeta r_0^2) r + \epsilon\, r_0^2 = \mathcal{O}(r^4)\, . 
\ee
The $\mathcal{O}(r^4)$ terms in \eqref{desarr} can be neglected if the non-zero 
solutions of the resulting cubic equation are such that $r^4\ll Q^2/T$, which is equivalent to 
\be\label{gm}
r \ll \sqrt{[\sqrt{G}Q]\, \ell_{gB}}\, . 
\ee
In other words, non-zero positive solutions to the cubic equation yield a good approximation to  Killing horizon radii when these radii are much less than the geometric mean of the classical characteristic lengths $\sqrt{G}Q$ and $\ell_{gB}$ associated to $Q$ and $T$ respectively. When this condition is satisfied \eqref{cubic} reduces, approximately, to 
\be\label{cubic2}
z^3 - \zeta z + a =0 \, , \qquad z= \frac{r}{r_0}\, , \quad a = \frac{\epsilon}{r_0}\, . 
\ee

For RN-like black holes with $r_+ \gg r_-$ it may happen that only the Cauchy-horizon radius $r_-$ (and not the event-horizon radius $r_+$) is sufficiently small to be given by a solution of \eqref{cubic2}. This occurs when $a/\zeta$ is sufficiently small and positive; we can then neglect the  $z^3$ term and the one solution is $z\approx a/\zeta$, or 
\be
r_- \approx \frac{\left(\EE_{\rm em}- M\right)}{b\sqrt{T}\left(Q-Q_{\rm cr}\right)}\, . 
\ee
This result confirms that $r_-\to 0$ when $M \to \EE_{\rm em}$ from below.  However, as we shall now explain, the full cubic equation \eqref{cubic2} provides a reliable picture of all three `phases' (S-like, RN-like, timelike naked singularity) in a neighbourhood of the tri-critical point where they meet ($\epsilon= \zeta=0$). This neighbourhood is restricted only by the requirement that $r_+=r_h$ be small enough (when an event horizon exists) to be a solution of the cubic equation. 

The cubic function of $z$ in \eqref{cubic} is a ``depressed cubic'': the sum of all three solutions is zero. The cubic's discriminant $\Delta$ is 
\be
\Delta = 4\zeta^3 -27a^2\, . 
\ee
The sign of $\Delta$ is significant, as follows:
\begin{itemize}

\item $\Delta >0$. There are three real roots, $r=(r_1,r_2,r_3)$ in terms of the radii, but as they sum to zero  the number of positive roots is either $1$ (event horizon) or $2$ (event horizon {\sl and} an interior Cauchy horizon). 

\item $\Delta <0$.  There is one real root. The other two are complex but their sum is real. The number of positive real roots is either $0$  (naked singularity) or $1$ (event horizon); there cannot be a Cauchy horizon. Notice that $\Delta<0$ whenever $\zeta<0$, which is equivalent to 
\be
Q < Q_{\rm cr} \, . 
\ee
Thus, any `small-charge' BI-type black hole is S-like. This is consistent with the conclusion in the subsection \ref{subsec:ECBH} that all extreme BI-type black holes have $Q>Q_{\rm cr}$. 

\item  $\Delta =0$. In this case at least two of the three roots coincide, but since $\Delta=0$ is possible only when $\zeta\ge0$ (with equality only at the triple point where $\zeta=\epsilon=0$), 
the two coinciding roots must be $r=r_+$ and $r=r_-$; i.e. $r_\pm =r_h$ is the radius of the degenerate horizon of an extreme black hole. Thus. $\Delta=0$ iff $M= M_{\rm ext}$, except at the triple point where $M= \EE_{\rm em}$ (since there is no extreme black hole at this point). 

\end{itemize}

All of the above is for the BI-type charged black hole holes ($\nu=\frac12$). We now turn to the 
$\frac34>\nu>\frac12$ cases. We shall focus on black-hole solutions for the `minimal' self-dual NLED family defined by the Hamiltonian CH-function of \eqref{frakh1}. The small-$r$ expansion of the $\EE(r)$ function for these cases is given in \eqref{EEsmall-r}. Using this expansion one finds the 
following Killing horizon equation near $r=0$:
\be
0= -\epsilon - A r^{3-4\nu} + r + B r^3 + \mathcal{O}\left(r^{7-4\nu}\right) 
\ee
where 
\be
A= \left(\frac{8\pi GT}{3-4\nu}\right) \left(\frac{Q^2}{2\nu T}\right)^\nu >0\, ,  \qquad 
B= \frac{8\pi GT}{3}\ \sim \ 1/\ell_{gB}^2\, . 
\ee 
For $r \ll \ell_{gB}$ we can neglect the $\mathcal{O}\left(r^{7-4\nu}\right)$ terms; we then have the following approximate equation 
\be\label{truncated}
r+ \epsilon = A r^{3-4\nu} - Br^3 \, ,  
\ee
where $r$ is restricted to be non-negative. 
The graph of the left-hand side is a straight line of unit slope and intercept $\epsilon$. The graph of the right-hand side is a concave curve that starts at zero (for $r=0$), rises monotonically to a maximum value of $r$ before falling monotonically, reaching zero again when 
\be
r^{4\nu} = \frac{A}{B}  \quad \Rightarrow \quad  r \ \sim \ \sqrt{(\sqrt{G} Q) \ell_{gB}} \, . 
\ee
This value of $r$ is much less than $\ell_{gB}$ when $\sqrt{G} Q \ll \ell_{gB}$, which we now assume.  Under these conditions any intersection of the two graphs for $r>0$ is the radius of a Killing horizon, and the number of these intersections cannot exceed $2$ (in agreement with the general theorem). The precise number ($0,1,2$) depends on the sign of $\epsilon$ as follows:
\begin{itemize}

    \item $\epsilon \le 0 $ ($M> \EE_{\rm em}$). As the intercept of the straight line is negative it intersects the concave curve at most once. As we know that $M\ge \EE_{\rm em}$ is possible only for 
    S-like black holes, we conclude that the intersection point is the event-horizon of an S-like black hole 

    \item $\epsilon = 0$ ($M =\EE_{\rm em}$). The straight line now intersects the concave curve twice but one intersection is at $r=0$, so there is still one Killing horizon, which must be (by continuity) the event horizon of an S-type black hole.

    \item $\epsilon > 0$ ($M < \EE_{\rm em}$). In this case the straight line will intersect the concave curve twice provided that $\EE_{\rm em} -M$ is sufficiently small, once at $r=r_-$ (near $r=0$) and 
    then again at $r=r_+$, which is still the radius of the event horizon but now of an RN-like black 
    hole. As $M$ is decreased the value of  $\EE_{\rm em} -M$  will increase until $r_+-r_-\to0$. We then have an extreme RN black hole that becomes a timelike naked singularity when $M$ is further decreased
    to $M< M_{\rm ext}$
\end{itemize}

To summarise: for sufficiently small $Q$ the equation \eqref{truncated} provides a reliable  approximation to the phase diagram near $r=0$; this is again a point where the three phases meet, but now they meet at $Q=0$ (rather than at some larger `critical' value, as in the BI-type case).

\subsection{Phase diagrams}\label{subsec:phased}

In this section we present full phase-diagram for the various ``types'' of charged black holes, using the parameter-space coordinates $(m,Q)$ where $m$ is the parameter proportional to $M/Q$ defined in \eqref{M/Q}. Recall that for RN-like black holes there is a minimum value $\mu$ of $m$ corresponding to the minimum mass $M_{\rm ext}$ for RN-like black holes, and that $\mu$ is a monotonically increasing function of $Q$ that approaches $\mu=1$ as $Q\to\infty$. 

One of the constraints of causality is that there are four `types' of spherically-symmetric black holes, corresponding to four qualitatively different phase diagrams, 
They differ according to which phases occur and the topological features of the boundary lines that separate them. One major difference is whether the electromagnetic energy is infinite or finite, and we consider separately these two possibilities.  

\begin{figure}[h!]
 \centering
\includegraphics[width=0.7\textwidth]{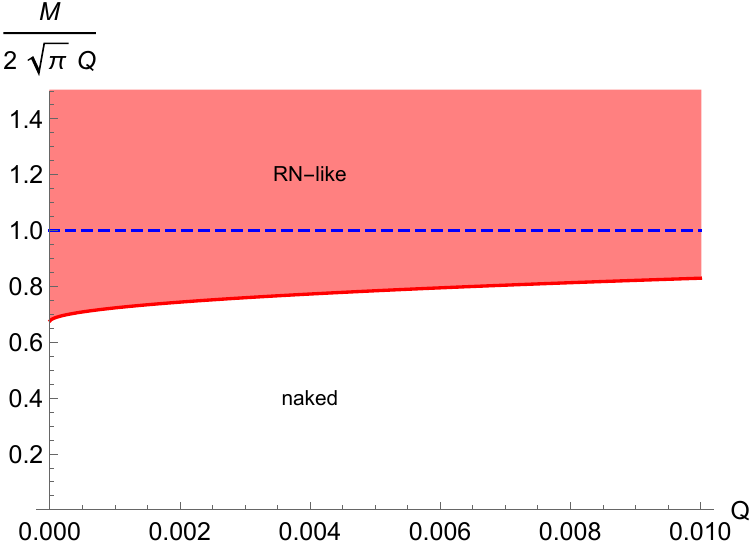}
  \caption
 {Phase diagram for a charged black hole in an interacting theory with $\nu=1$. $M= M_{\rm ext}$ on the red boundary curve $m\equiv \sqrt{\frac{G}{4\pi}}\, \frac{M}{Q}=\mu(Q)$, which is asymptotic to the horizontal dashed blue line: $m= \mu^{\rm RN} \equiv 1$. We have set $G=1$.}
 \label{fase11}
 \end{figure}
 
\subsubsection{Infinite electromagnetic energy}

A feature of infinite electromagnetic energy is that there are only two phases, 
RN-like black holes and naked singularities, separated by the $M=M_{\rm ext}$ curve
in parameter space. These cases are relatively simple but there are novelties, and two ``types'' with `phase-diagrams that have not previously been discussed.  In terms of the $(m,Q)$ coordinates, the $M=M_{\rm ext}$ curve is $m=\mu(Q)$. The two types of black hole differ in how the curve $\mu(Q)$ behaves as 
$Q\to0$:
\bigskip

\noindent {\bf Maxwell-type}: {$\nu=1$}. 
   Because Maxwell is a free field theory, the Hamiltonian function for RN black holes has the same $1/r^4$ behaviour for both large and small $r$, and $\mu(Q)\equiv 1$. 
   However, there are {\sl interacting} theories for which ${\rm H}\sim 1/r^4$ for both large and small $r$, but with different coefficients. In these cases the curve $m=\mu(Q)$ increases monotonically to $\mu=1$ from some initial value $0<\mu_0<1$.  The typical phase diagram is illustrated in fig. \ref{fase11}.
   
   An explicit example is provided by the self-dual theory with the Hamiltonian CH-function 
\be
\frak{h}(\sigma) =\sigma +T- T\sqrt{1+\frac{a^2\, \sigma^2}{T^2}}\ , 
\ee
with $0<a\le 1/\sqrt{2}$ needed to satisfy the causality conditions \eqref{hfrak-c}. In the weak field limit, $\frak{h}(\sigma)\approx \sigma$, and in the strong field limit,
$\frak{h}(\sigma)\approx (1-a)\sigma$. The corresponding Hamiltonian function defines a family of causal NLED theories with Maxwell as the weak-field limit and ModMax as the strong-field limit. For $p=0$, the ModMax Hamiltonian is 
\be
H_{MM}(s,0) := H(s) = e^{-\gamma} s\, ,  \qquad e^{-\gamma} := 1- a \, ,  
\ee
which yields $\mu_0= \sqrt{1-a}$. 
\bigskip

   \begin{figure}[h!]
 \centering
\includegraphics[width=0.7\textwidth]{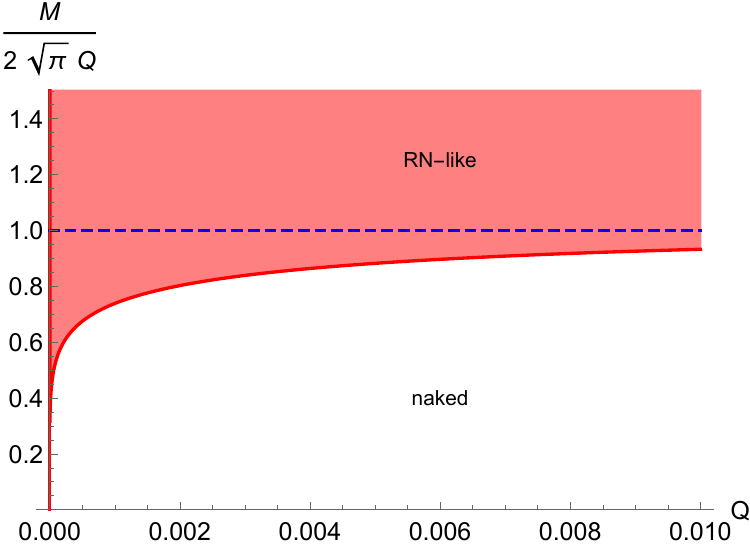}
  \caption
 {Phase diagram for a charged black hole in a theory with $\nu=\frac78$.  The red boundary curve $m=\mu(Q)$, is asymptotic to the horizontal dashed blue line corresponding to extreme RN black holes. Its slope becomes infinite as $Q\to0$.  }
 \label{fase78}
 \end{figure}

    \noindent {\bf Intermediate-type I}: {$\frac34 \le \nu <1$}.  The curve $m=\mu(Q)$ representing extremal black holes is now one for which $\mu(Q) \to 0$ as $Q\to 0$. 
    If Maxwell is the weak-field limit then $\mu(Q) \to 1$ as $Q\to\infty$; if ModMax
    is the weak-field limit then $\mu\to e^{-\gamma/2}Q$ as $Q\to\infty$.  

    This type of phase diagram is illustrated in fig. \ref{fase78}. Notice that the slope of the curve $m=\mu(Q)$ diverges at $Q=0$. This is a general feature which can be proved using the small-$r$ expansion of $\mu(Q)$ near $Q=0$.

\subsubsection{Finite electromagnetic energy}

In these cases there are three ``phases'' because we now have S-like black holes for $M\ge \EE_{\rm em}$. This gives us the S-like boundary line $M= \EE_{\rm em}$.
In addition, we still have the lower boundary line for RN-like black holes 
$M= M_{\rm ext}$. As explained in subsection \ref{subsec:ECBH}, we now get two types 
according to whether the two boundary curves meet at $Q=0$ or at $Q=Q_{\rm cr}>0$. 
We start with the first of these, which has not  previously been discussed. 
\bigskip 

\noindent {\bf Intermediate-type II}: $\frac12 <\nu<\frac34$.
The curve $m= \mu(Q)$ in the $(m,Q)$-plane lies below the S-like boundary curve except at $Q=0$ where the two curves meet)
and, as for Intermediate-type I black holes, this curve has the property that its slope at $Q=0$ is infinite.

\begin{figure}[h!]
 \centering
\includegraphics[width=0.7\textwidth]{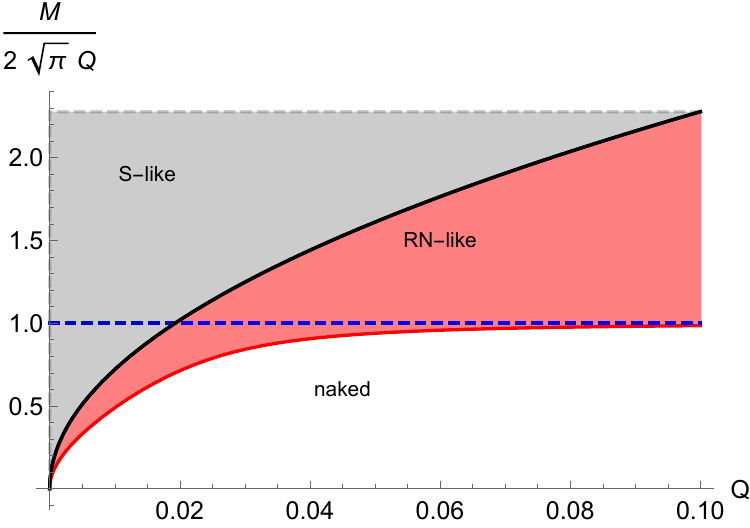}
  \caption
 {Phase diagram for a charged black hole in a theory with $\nu=\frac58$. The (upper) black curve is  $M= \EE_{\rm em}$, and the (lower) red curve is $M= M_{\rm ext}$.}
 \label{fase58}
 \end{figure}

A simple example is provided by charged black-hole solutions of the family of self-dual NLED theories defined by the Hamiltonian CH-function of \eqref{frakh1}, for a choice of $\nu$ in the appropriate range ($\frac12 <\nu<\frac34$). Using the small-$r$ expansion of 
\eqref{EEsmall-r} we find that 
\be 
\label{gyp}
-g_{tt}=  -\frac{2G(M-\EE_{\rm em})}{r}
-\frac{2c_\nu G T^{1-\nu} Q^{2\nu}}{r^{4\nu-2}}+1  +\cdots
\ee
where $c_\nu$ is the positive numerical factor given in \eqref{Enu}.
It should be noted that this result differs from that assumed in \cite{Hale:2025ezt} because (as we remarked in subsection \ref{subsec:bh-sd}) their assumed form of $\EE(r)$ was essentially that of the $\nu=\frac12$ case.\footnote{The function $\EE(r)$ was called $U_{\rm self}(r)$ in \cite{Hale:2025ezt}.}

Whereas the second term on the right-hand side of \eqref{gyp} is a constant for $\nu=\frac12$ (which we consider below as the fourth `type') it blows up as  $1/r^k$, with $0<k<1$ for $\frac12 <\nu<\frac34$. This difference allows RN-like black holes to exist for all $Q>0$. The phase diagram is illustrated in 
fig. \ref{fase58} for $\nu=\frac58$.

\begin{figure}[h!]
 \centering
\includegraphics[width=0.7\textwidth]{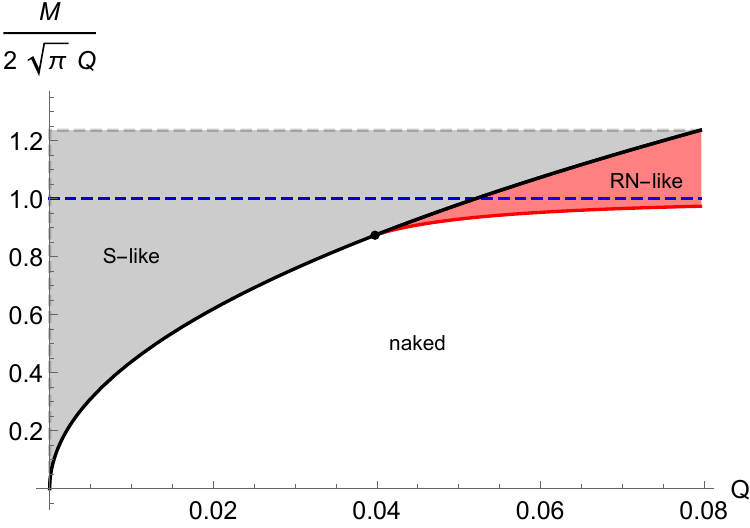}
  \caption
 {Phase diagram for a BI black hole, illustrating BI-type. The three phases met at a triple point, where $Q = Q_{\rm cr}$. The black curve is $M= \EE_{\rm em}(Q)$;  
 for $Q<Q_{\rm cr}$ it represents a zero horizon-radius limit of S-like black holes.  
 The red curve for $Q > Q_{\rm cr}$ is the $m= \mu(Q)$ curve, which is asymptotic to the horizontal blue line and ends at the triple point.}
 \label{fased}
 \end{figure}

\bigskip

\noindent {\bf Born-Infeld type}: $\nu=\frac12$. The phase diagram for this case was found in \cite{Hale:2025ezt} but presented in the $(Q/M,M)$-plane, rather than the $(M/Q,Q)$-plane used here. 
The RN-like lower boundary line now ends on the S-like lower boundary line at $Q= Q_{\rm cr}$. This is a triple point, where the radius of the extreme black hole shrinks to zero. There is now a minimum value of 
$\mu$, its value at the triple point:
\be
\mu_{\rm tri} = \frac{ G\EE_{\rm em}(Q_{\rm cr})}{\sqrt{4\pi G}\, Q_{\rm cr}} \, . 
\ee
This can be computed explicitly for the family of self-dual BI-type NLED theories defined for parameter 
$\alpha\in (0,1)$ by the Lagrangian CH-function of \eqref{arem}. The critical charge is
\be
Q^{(\alpha)}_{\rm cr}=\frac{1}{8\pi G\sqrt{2\alpha T}}\ ,  
\ee
which yields the BI result of \eqref{Qcrit} for $\alpha=\frac12$ (as it must since the $\alpha=\frac12$ case {\sl is} BI). Using this in the expression of \eqref{energyalfa} for $\EE_{\rm em}$ we find that
\be
\mu^{(\alpha)}_{\rm tri} =\frac{\pi  \Gamma
   \left(\frac{3}{2}-\frac{\alpha }{2}\right)}{4 \Gamma
   \left(\frac{7}{4}\right)
   \Gamma
   \left(\frac{7}{4}-\frac{\alpha }{2}\right)}\ \in \ (0.824,0.943)\, .  
   \ee
This is a monotonically increasing function of $\alpha$; the range given is an approximate range as  $\alpha$ increases from $0$ to $1$. For the BI case we set $\alpha=\frac12$ to get
\be 
\label{mucrBI}
\mu^{\rm BI}_{\rm tri} = \frac{\Gamma\Big(\frac14\Big)^2}{6\sqrt{2\pi}}\approx 0.874\  .
\ee

\subsection{Geometry of the Born particle}\label{subsec:BV}

In addition to the three phases for finite electromagnetic energy there are also three kinds of boundary separating them. One is standard because it is a feature of the phase diagram for the RN family of spacetimes; it is the $M= M_{\rm ext}$ curve separating RN-like black holes from timelike naked singularities. A second kind is generic for finite electromagnetic energy; this is the boundary between RN-like black holes and S-like black holes. Both kinds can be understood from the RN-like side as limits in which the Cauchy horizon disappears by moving to coincide with either the event horizon or the singularity. The third kind occurs only for BI-type black holes; this is the ``small charge'' boundary between S-like black holes and naked singularities; it can be understood from the S-like side as a limit of zero event horizon radius, but without the prior transition to an RN-type black hole. 

As both the second and third kinds of phase boundary are segments of the $M= \EE_{\rm em}$ curve in the phase diagram, we can consider them together:
\begin{itemize}

    \item $Q>Q_{\rm cr}$. This is the boundary between RN-like and S-like black holes. From outside the event horizon this transition is not obviously apparent.

    \item  $Q<Q_{\rm cr}$. This is the boundary between S-like black holes and naked singularities. The boundary itself corresponds to naked singularities, but of an interesting kind. Firstly, they 
    are obtainable from a zero entropy limit of S-like black holes, they are potentially elementary pointlike charged particles. Secondly, they have a mass equal to their electromagnetic energy, thus 
    realising the idea by Born that motivated his original NLED theory \cite{Born:1933qff}. Remarkably, it seems that such a ``Born particle'' requires not only an interacting electrodynamics theory but also gravity. 
    
\end{itemize}
For both cases we get the spacetime metric near $r=0$ by setting $M=\EE_{\rm em}$ (equivalently, $\epsilon=0$) into the formula of \eqref{desarr}. This yields 
\be 
-g_{tt} \equiv \NN^2 \, = \,   -\zeta + \frac{r^2}{r_0^2} + \mathcal{O}(r^4)\, , 
\ee
where, from \eqref{ep} and \eqref{zet}, 
\be
\label{zeet}
\zeta =-\left(1-\frac{Q}{Q_{\rm cr}}\right)\  \sim \ \frac{\sqrt{G}(Q- Q_{\rm cr})}{\ell_{gB}}\, , \qquad r_0\ \sim\  \ell_{gB}\, . 
\ee 
Away from the triple point we have, as $r\to0$, 
\be
ds^2 \ \sim\  \zeta dt^2 - \zeta^{-1} dr^2 + r^2d\Omega^2 \, . 
\ee 

For $\zeta>0$, equivalently $Q>Q_{\rm cr}$, the time coordinate is $r$ rather than $t$, and hypersurfaces of constant $r$ are spacelike hypersurfaces of topology $\bb{R}\times S^2$ with the 2-sphere collapsing to a point at $r=0$. However, this singular hypersurface is not a naked singularity because there is an event horizon for  $Q>Q_{\rm cr}$ and $t$ is the time coordinate outside this horizon. 

For $\zeta<0$, equivalently $Q<Q_{\rm cr}$, the spacetime metric as $r\to0$ is 
\be\label{BVm}
ds^2 \ \sim\  - d\tau^2  + d\rho ^2 + |\zeta| \rho^2d\Omega^2 \, . 
\ee 
where $(\tau,\rho)$ are the following rescaled time and radial coordinates:
\be
\tau= \sqrt{|\zeta|}\, t \, , \qquad \rho = r/\sqrt{|\zeta|}\, . 
\ee
An important point to appreciate here is that, from the definition of $\zeta$ in \eqref{zeet} and for $0\leq Q< Q_{\rm cr}$, we have
\be
0\leq (-\zeta) < 1\, .
\ee
In the $Q=0$ case the 4-metric of \eqref{BVm} is the metric for Minkowski 
space in spherical polar coordinates. Since $Q\ne0$ for us, we have the strict inequality $(-\zeta) < 1$. In these cases the 4-metric \eqref{BVm} is a direct product of $\bb{R}$ (time) with a ``non-flat hypercone''. It is a hypercone because the solid angle subtended by a 2-surface of constant $\rho$ is 
$4\pi |\zeta|$ rather than $4\pi$, implying a deficit angle of $4\pi(1-|\zeta|)$.
It is ``non-flat'' because the Ricci curvature scalar is non-zero, and finite for any $\rho>0$, but blows up as $\rho^{-2}$ as $\rho\to 0$; in the unscaled coordinates one finds that, as $r\to0$, 
\be
R\ \sim \  \frac{2Q}{Q_{\rm cr}\ r^2}\ .
\ee
As expected, this is zero when $Q=0$.  

Remarkably, the metric \eqref{BVm} has arisen previously as the Barriola-Vilenkin metric describing a ``global monopole'', with $(1-|\zeta|)/(8\pi G)\propto \sqrt{T}Q$ being the topological monopole charge \cite{Barriola:1989hx,Tan:2017egu}. In the current context, the tip of the hypercone is 
a ``Born particle'', i.e. a charged particle of mass $M^{\rm Born}= \EE_{\rm em}$, which is restricted by the ``small-charge'' condition $Q<Q_{\rm cr}$ since this implies
a maximum mass; approximately, 
\be
GM^{\rm Born} < \ell_{gB}\, . 
\ee
The mass to charge ratio of a Born particle is less 
than it is for any extreme black hole because the mass at the triple point is a lower bound on extreme black hole mass.
In fact, it can be arbitrarily small.
This follows from
the expression of \eqref{S-boundary} for $\EE_{\rm em}$, which implies (again approximately) that
\be
\left(\frac{GM^{\rm Born}}{\sqrt{G}{Q}}\right)^2 \sim \frac{\sqrt{G}Q}{\ell_{gB}} \, . 
\ee

\section{Summary and Discussion}

In the context of General Relativity, nonlinear theories of electrodynamics allow a natural generalisation of the Einstein-Maxwell equations, and hence of the Reissner-Nordstr\"om (RN) charged black-hole solutions. Although this generalisation  introduces a new classical length scale via the ``gravitational-Born length'' $\ell_{gB} \sim 1/\sqrt{GT}$ associated to the Born tension $T$, it does so without changing the basic character of the equations (second-order PDE) that one expects of a semi-classical approximation to a quantum theory in which quantum effects become important at the Planck length. 

However, the propagation of superluminal signals in some backgrounds is still possible, so causality conditions on the NLED Lagrangian or Hamiltonian are necessary. This fact has been appreciated since the late 1960s,  but the necessary and sufficient conditions were found relatively recently \cite{Schellstede:2016zue}. We pointed out in 
\cite{Russo:2024kto} that they can be divided
into weak-field causality conditions (which could be violated for arbitrarily weak fields) and 
one strong-field causality condition (which requires some fields to be strong relative to the energy density scale set by the Born tension); this distinction can also be understood in other ways \cite{Russo:2024xnh,Russo:2025fuc}. 

We have argued, using a class of NLED
theories that satisfy only the weak-field causality conditions, that a violation of the strong-field causality condition leads to an instability 
of black-hole solutions of the Einstein-NLED equations. Our main aim, however, has been to explore the restrictions on spherically symmetric black-hole solutions of the Einstein-NLED equations implied purely by causality of the NLED theory.  

One implication of strong-field causality is that the NLED stress-energy tensor must satisfy the Strong Energy Condition (SEC) \cite{Russo:2024xnh}. As we pointed out in that work, this fact, taken together with one of the Hawking-Penrose singularity theorems \cite{Hawking:1970zqf}, implies a spacetime singularity; i.e. the absence of ``regular'' black-hole solutions of the Einstein-NLED equations for any causal NLED theory with a weak-field expansion. Here we have used more direct methods to confirm this conclusion. 

Other implications of causality for Einstein-NLED black holes have been investigated previously, and we have both clarified and extended these results in various ways. One difference from earlier works is our use of the Hamiltonian formulation of NLED theories. In the context of spherically-symmetric charged black holes the Hamiltonian function becomes a ``black-hole'' function ${\rm H}(r)$ that determines another function $\EE(r)$ equal to the electromagnetic energy outside a ball of radius $r$. It was shown in \cite{Hale:2025ezt} that $\EE(r)$ is a convex function whenever the NLED stress-energy tensor satisfies the SEC, and this fact was used to deduce causality constraints on the global structure of charged black hole spacetimes. 

For reasons summarised below, not all of the results of \cite{Hale:2025ezt} are valid for {\sl all} causal NLED theories. However, their conclusion (for asymptotically-flat and spherically-symmetric spacetimes) that there can be no more than two Killing horizons is generally valid, but we have shown that this result actually depends on {\sl strict} convexity of $\EE(r)$ for finite $r$, which is a consequence of causality that goes beyond the SEC.  

We have also provided simplifications and clarifications of other previous results, particularly those of \cite{Abe:2025vdj} on the monotonicity in the charge $Q$ of various functions of relevance to issues 
dealt with here, and to black-hole thermodynamics \cite{Abe:2025vdj} which we do not discuss here. For example, our use of the Hamiltonian formulation allows consideration of all charge ratios $q_e/q_m$ at once. Also, our derivation of a monotonicity condition of \cite{Abe:2025vdj} for $\mu(Q)$ (the extremal black-hole mass-to-charge ratio) makes it clear that this is a consequence of weak-field causality only. In contrast, both weak-field and strong-field causality conditions are needed to show that the function $r_h(Q)/Q$ is a monotonic increasing function of $Q$, where $r_h$ is the event-horizon radius of an extreme black hole. This implies
(in the quantum theory) that the ratio of the entropy to $Q^2$ is a monotonic increasing function of $Q$ at fixed $M$. Since $Q\to\infty$ is a free-field limit for fixed $M$, we conclude that causal NLED interactions reduce the entropy of an extreme black hole at fixed $Q$, with a reduction that increases with the strength of the interactions. 

Any spherically-symmetric solution of the Einstein-NLED equations that is asymptotic to a Reissner-Nordstr\"om spacetime for large $r$ has the mass $M$ and charge $Q$ as independent parameters, possibly in addition to other dimensionless parameters such as the charge ratio $q_e/q_m$. For a given NLED theory and charge ratio, 
the solution for any choice of $(M,Q)$ will be either a spacetime with a naked singularity or a black hole, which may be RN-like or S-like (without a Cauchy horizon \cite{deOliveira:1994in}). We have represented these possibilities as ``phase-diagrams'' in the $(m,Q)$-plane, where $m$ is a mass-to-charge ratio normalized such that $m=1$ for an extreme RN black hole. We have found that there are four ``types'' of qualitatively different phase diagrams, corresponding to four ``types'' of black hole; more precisely, charged spherically symmetric spacetime families parameterised by $(m,Q)$. 

A useful guide to understanding the distinction between the various ``types'' is the assumption that the black-hole function ${\rm H}(r)$ behaves as a power of $r$ near $r=0$:
\be
r\to 0 \ : \quad {\rm H}(r) \sim r^{-4\nu} \,  .
\ee
For example, this assumption can be used to determine the behaviour of $\EE(r)$ near $r=0$, and strict causality of $\EE(r)$ in this region requires $\nu\ge\frac12$. We get an upper-bound on $\nu$ by using the fact that $\nu\le1$ is required for concavity of the ``effective charge-squared function'' 
\be
Q^2_{\rm eff}(r) = \frac{r}{2\pi} \EE(r) \, .  
\ee
This concavity condition is equivalent to non-positivity of the function $\Theta(r)$ (the trace of the stress-energy tensor as a function of $r$), and is therefore a weak-field causality condition. Combining these results we see that causality requires
\be\label{range}
\tfrac12 \le \nu \le1 \, . 
\ee
For $\nu=1$ we have Maxwell, but also some novel interacting causal NLED theories with ModMax as a strong-field limit. For $\nu=\frac12$ we have Born-Infeld.  In fact, for every choice of $\nu$ in the range of \eqref{range} there is a simple explicit causal interacting self-dual NLED theory. However, different values of the power-index $\nu$ do not necessarily lead to qualitatively different phase diagrams, and we associate the four different ``types'' with values, or ranges of values, of $\nu$. The values of $\nu$ separating these regions, which include $\nu=\frac12,1$ are therefore special. 

 The only other special value is $\nu=\frac34$ because it separates the black holes with finite electromagnetic energy ($\nu<\frac34$) from those with infinite electromagnetic energy ($\nu\ge\frac34$). We remark, in passing, that infinite electromagnetic energy is consistent with finite mass because the gravitational energy may be negative and infinite; the physical mass is the finite and positive ADM mass $M$. Nevertheless, the distinction is important because S-like black holes are possible only if the electromagnetic energy is finite \cite{Hale:2025ezt}. 

 Notice that the special values of $\nu$ are those for which $4\nu=2,3,4$. By an analysis of the behaviour as $r\to 0$ of $r^n{\rm H}(r)$ for $n=2,3,4$, we have shown the distinctions 
 between the four ``types'' of black hole arrived at via the power-law assumption for ${\rm H}(r)$
 as $r\to0$ remain valid even when the power-law assumption defining $\nu$ fails. 

The most-studied of the four ``types'' is what we call the ``BI-type''  because it includes the black-hole solutions of the Einstein-Born-Infeld equations; the ``phase-diagram'' for this type is the one presented in \cite{Hale:2025ezt} (in a different parameterisation to ours). It has the  feature that all black holes are S-like for sufficiently small charge $Q$ (and there is some evidence that an analogous feature persists in the rotating case \cite{Hale:2025urg}). This property, discovered in \cite{deOliveira:1994in},
was claimed in \cite{Hale:2025ezt} to be a feature of all black-hole solutions of causal NLED theories for which the electromagnetic energy is finite, but it is actually a property only of BI-type black holes. 

Charged black holes with $\frac12 <\nu<\frac34$ are another ``type'' with finite electromagnetic energy but without a critical charge (equivalently, a critical charge equal to zero). Crucially, the small-$r$ expansion of $\EE(r)$ for this type differs from the BI-type, which was simply assumed in \cite{Hale:2025ezt}.

For infinite electromagnetic energy there are also two ``types''. One we call Maxwell-type ($\nu=1$); it includes the RN black hole solutions of the Einstein-Maxwell equations, but also a novel class of interacting causal NLED theories with ModMax as a strong-field limit. The other ``type'' corresponds to $\frac34<\nu<1$,
and is disinguished by a very different behaviour of the extreme mass-to-charge ratio $\mu(Q)$ as $Q\to0$.

One important point concerning this classification into ``types'' is that we are not classifying causal NLED theories, except within the special class of self-dual theories. This is because a 
{\sl non-self-dual} causal NLED theory may allow black holes with different values of $\nu$, according to the $q_e/q_m$ charge ratio. We have given an explicit example that illustrates this point; its electric and magnetic black holes have different values of $\nu$, and the dependence of $\nu$ on the $q_e/q_m$ charge ratio is not  everywhere continuous. 

One of the main novelties of Einstein-NLED black holes is the possibility of $S$-like black holes, and one 
interesting aspect of them concerns the electric and magnetic flux lines of the divergence-free vector density fields $(D,B)$ on a constant-$t$ spacelike hypersurface. Recalling that the complete Schwarzschild spacetime has four regions, and that regions  I and IV are connected by an Einstein-Rosen bridge across a minimum two-sphere at the horizon radius, we see that the flux lines must pass smoothly through region I into region IV, where they presumably expand out to another spatial infinity. This appears to be an example of ``charge without charge'' \cite{Misner:1957mt}. Evaporation of an S-like black hole due to Hawking radiation must eventually cause the Einstein-Rosen bridge to collapse to a point. Generically, this will happen at a transition (behind an event horizon) from S-like to RN-like black hole; the reverse transition is one in which a Cauchy horizon shrinks to $r=0$. 

However, for ``small-charge'' BI-type black holes (with $Q < Q_{\rm cr}$) only S-like black holes exist, and Hawking evaporation can be expected to reduce the mass $M$ until $M=\EE_{\rm em}$. At this point 
it has become a ``Born particle'': a charged particle with a mass equal to the energy of its electromagnetic fields. We have shown that the spacetime geometry of a Born particle is
that of the Bariola-Vilenkin global monopole. The centre of symmetry is a naked ``non-flat hyperconical'' singularity; with the charge proportional to the solid angle deficit. 


\section*{Acknowledgements}

JGR acknowledges financial support from the Spanish  MCIN/AEI/10.13039/501100011033 grant PID2022-126224NB-C21.



\end{document}